\documentclass[a4paper,fleqn,usenatbib]{mnras}

\usepackage{newtxtext,newtxmath}

\usepackage[T1]{fontenc}
\usepackage{ae,aecompl}

\usepackage{graphicx}	
\usepackage{amsmath}	
\usepackage{amssymb}	

\title[Environment-dependent galaxy evolution since $z \sim 3$]{Reconstructing the galaxy density field with photometric redshifts: II. Environment-dependent galaxy evolution since $z \simeq 3$}

\author[N. Malavasi et al.]{
Nicola Malavasi,$^{1,2}$\thanks{E-mail: nmalavas@purdue.edu}
Lucia Pozzetti,$^{3}$
Olga Cucciati,$^{1,3}$
Sandro Bardelli,$^{3}$
\newauthor
Olivier Ilbert,$^{4}$
and Andrea Cimatti$^{1,5}$
\\
$^{1}$Universit\`{a} di Bologna -- Dipartimento di Fisica e Astronomia (DIFA), v.le Berti Pichat 6/2 - 40127 Bologna, Italia\\
$^{2}$Department of Physics and Astronomy, Purdue University, 525 Northwestern Avenue, West Lafayette, IN 47907, USA\\
$^{3}$INAF -- Osservatorio Astronomico di Bologna, via Ranzani 1 - 40127 Bologna, Italia\\
$^{4}$Aix Marseille Universit\'{e}, CNRS, LAM (Laboratoire d'Astrophysique de Marseille) UMR 7326, 13388, Marseille, France\\
$^{5}$INAF -- Osservatorio Astrofisico di Arcetri, Largo E. Fermi 5 - I-50125 Firenze, Italia\\}

\date{Accepted 2017 May 25. Received 2017 May 23; in original form 2016 August 31}

\pubyear{2017}

\begin{document}
\label{firstpage}
\pagerange{\pageref{firstpage}--\pageref{lastpage}}
\maketitle

\begin{abstract}
Although extensively investigated, the role of the environment in galaxy formation is still not well understood. In this context, the Galaxy Stellar Mass Function (GSMF) is a powerful tool to understand how environment relates to galaxy mass assembly and the quenching of star-formation. In this work, we make use of the high-precision photometric redshifts of the UltraVISTA Survey to study the GSMF in different environments up to $z \sim 3$, on physical scales from 0.3 to 2 Mpc, down to masses of $M \sim 10^{10} M_{\sun}$. We witness the appearance of environmental signatures for both quiescent and star-forming galaxies. We find that the shape of the GSMF of quiescent galaxies is different in high- and low-density environments up to $z \sim 2$ with the high-mass end ($M \gtrsim 10^{11} M_{\sun}$) being enhanced in high-density environments. On the contrary, for star-forming galaxies a difference between the GSMF in high- and low density environments is present for masses $M \lesssim 10^{11} M_{\sun}$. Star-forming galaxies in this mass range appear to be more frequent in low-density environments up to $z < 1.5$. Differences in the shape of the GSMF are not visible anymore at $z > 2$. Our results, in terms of general trends in the shape of the GSMF, are in agreement with a scenario in which galaxies are quenched when they enter hot gas-dominated massive haloes which are preferentially in high-density environments.
\end{abstract}




\section{Introduction}
\label{intro}
The current understanding of the galaxy formation and evolution paradigm strongly relies on observational evidence of a correlation between galaxy properties and the environment in which galaxies reside. Although the ways for defining galaxy environment are numerous, starting from early works on the morphology-density relation \citep[\textit{e.g.}][]{1980ApJ...236..351D}, evidence has been gathered on the existence of a relation between local density and galaxy properties, such as colour, star-formation, stellar mass, and size \citep[see \textit{e.g.}][]{2005ApJ...629..143B,2004MNRAS.353..713K,2006MNRAS.370..198C,2012MNRAS.419.3018C,1998ApJ...504L..75B,2004ApJ...615L.101B,2003ApJ...584..210G,2002MNRAS.334..673L,2006A&A...458...39C,2007A&A...468...33E}. There is general agreement over galaxies in high-density environments being more massive, less star-forming and generally more evolved in comparison to low-density environments \citep[see \textit{e.g.}][for a review on environmental properties of nearby galaxies]{2009ARA&A..47..159B}.

The role played by environment, defined both in terms of the local density field, as well as of global Large Scale Structure features (such as clusters, filaments and voids) is still poorly understood. High-density regions are characterized by specific processes (such as interactions of galaxies with the hot intracluster medium, interactions of galaxies with a cluster potential well or interactions of galaxies with other galaxies) which can easily interrupt the star-formation. Moreover, as galaxies are biased tracers of the underlying dark matter distribution, different galaxy samples and/or different environment parametrizations may probe different kinds of local and global environment (\textit{i.e.} DM haloes with different mass, see \textit{e.g.} \citealt{2012MNRAS.419.2670M,2015MNRAS.446.2582F, 2012MNRAS.419.2133H}). On the other hand, the galaxy stellar mass is related to the halo mass. It is also for this reason that it has not yet been determined whether star-formation quenching can be separated in two distinct processes (one depending only on environment and one on galaxy stellar mass, as proposed \textit{e.g.} by \citealt{2010ApJ...721..193P}) or whether mass and environment are just two aspects of the same underlying physical mechanisms (as proposed \textit{e.g.} by \citealt{2015MNRAS.447..374G}, see also Sect. \ref{discussion}).

One of the best ways to study galaxy evolution in different environments is to compare the Galaxy Stellar Mass Function (GSMF) as a function of redshift, galaxy type (star-forming or quiescent), and environmental density. GSMFs are a powerful tool as they allow to summarize in a single distribution function the galaxy number density as a function of mass and to study its evolution with redshift or its dependence on other galaxy properties such as colour and star-formation activity. The study of the shape of the GSMF is a powerful indicator of how the build-up of galaxy mass happens throughout cosmic history. Moreover, theory and numerical simulations can make predictions for the shape of the GSMF to be compared with observations and therefore understand the physical processes responsible for galaxy evolution. Many works have already studied the comparison of the predicted GSMF from semi-analytical models and simulations and the observed GSMF (especially in different environments, see \textit{e.g.} \citealt{2011MNRAS.413..101G, 2012MNRAS.422.2816B, 2006A&A...459..745F, 2009ApJ...701.1765M, 2009ApJ...707.1595D,2009MNRAS.397.1776F,2009MNRAS.399..827L,2010MNRAS.401.1166C,2014ApJ...788...57V,2010A&A...523A..13P,2010A&A...524A..76B}). These works find that in the case of galaxies in the general field, the number of low-mass galaxies with old stellar populations is over-predicted at intermediate redshifts ($z > 0.5$), while the number of high-mass galaxies ($M \gtrsim 10^{11} M_{\sun}$) is under-predicted at high redshift ($z > 2$). In particular, \citet{2014ApJ...788...57V} performed a comparison between model and observed GSMFs in different global environments, finding that the discrepancies at low masses are present also for the cluster GSMF. Moreover, the models fail to reproduce the observed evolution for the high-mass end of both the cluster and the field GSMF. A detailed comparison of the GSMF presented in this work with GSMFs in different environments derived from semi-analytical models of galaxy formation (included, but not limited to the one used in \citealt{2016A&A...585A.116M}) will be the subject of a future paper.

Several studies have addressed the investigation of the GSMF using spectroscopic redshift surveys, from the local Universe \citep[see \textit{e.g.}][]{2004ApJ...600..681B,2008MNRAS.388..945B,2012MNRAS.421..621B}, using surveys such as the Sloan Digital Sky Survey \citep[SDSS,][]{2000AJ....120.1579Y} or the Galaxy and Mass Assembly survey \citep[GAMA,][]{2011MNRAS.413..971D}, up to $z \sim 1$ \citep[see \textit{e.g}][]{2004A&A...424...23F,2007A&A...474..443P,2010A&A...523A..13P,2013ApJ...767...50M,2013A&A...558A..23D}, relying on data from surveys such as the K20 survey \citep{2002A&A...392..395C}, the VIMOS VLT Deep Survey \citep[VVDS,][]{2005A&A...439..845L}, the zCOSMOS survey \citep{2007ApJS..172...70L}, the PRIsm MUlti-object Survey \citep[PRIMUS,][]{2011ApJ...741....8C}, and the VIMOS Public Extragalactic Redshift Survey \citep[VIPERS,][]{2014A&A...566A.108G}. Photometric redshift surveys such as the COSMOS-UltraVISTA \citep{2007ApJS..172....1S,2012A&A...544A.156M}, and the VIPERS Multi-$\lambda$ Survey \citep[VIPERS-MLS][]{2016A&A...590A.102M} have instead been intensively used to explore the GSMF up to $z \sim 3$ \citep[see \textit{e.g.}][]{2010ApJ...709..644I,2013ApJ...777...18M,2013A&A...556A..55I,2016A&A...590A.103M}.

The GSMF in different environments has been thoroughly investigated in several works, again relying both on spectroscopic surveys of local galaxies (see \textit{e.g.} \citealt{2006ApJ...651..120B}, who used SDSS data, \citealt{2014MNRAS.445.2125M}, who used GAMA data, and \citealt{2001ApJ...557..117B}, who relied on the Two Micron All Sky Survey, 2MASS, \citealt{2000AJ....119.2498J}, and the Las Campanas Redshift Survey, LCRS, \citealt{1996ApJ...470..172S}). Spectroscopic surveys allowed the study of the GSMF also up to $z \sim 1 - 1.5$ \citep[see \textit{e.g.}][]{2003MNRAS.346....1K,2010A&A...524A..76B,2010MNRAS.409..337C,2016A&A...585A.160A,2012MNRAS.420.1481V,2016A&A...586A..23D,2015ApJ...806..162H,2012A&A...538A.104G}, while photometric redshift surveys have been used up to $z \sim 3$ \citep[see \textit{e.g.}][]{2013ApJS..206....3S,2015ApJ...805..121D,2015MNRAS.447....2M}. For example, \citet{2010A&A...524A..76B} studied the GSMF in different environments in the COSMOS field \citep[using the zCOSMOS survey, see][]{2007ApJS..172...70L,2009ApJS..184..218L} up to $z = 1$, finding a difference between the GSMF of high- and low-density environments, with the massive end of the GSMF being more enhanced in high-density environments. This result has been confirmed also by \citet{2016A&A...586A..23D}, by means of the VIPERS survey \citep{2014A&A...566A.108G,2014A&A...562A..23G}. Both these works relied on a local measurement of the environment, which is the same strategy adopted in this work. A complementary approach often used at $z \le 1-1.5$ is the study of the GSMF in different global environments, \textit{e.g.} by comparing the GSMF in clusters and in the field. This approach is substantially different from ours and for this reason it can yield very different results \citep[see \textit{e.g.}][]{2013A&A...557A..15V,2013A&A...550A..58V,2011MNRAS.412..246V,2016A&A...592A.161N,2013MNRAS.432.3141C}. In all these works, the GSMF does not seem to depend on global environment, the variations being small \citep{2013MNRAS.432.3141C}.

The difficulty in performing environmental studies at high redshift relies mainly in the scarce availability of spectroscopic surveys of all galaxy types that sample a large enough volume (wide area and deep limiting magnitude) with large enough statistical samples. Using photometric redshift surveys performed in the COSMOS field \citep{2012A&A...544A.156M,2013A&A...556A..55I}, \citet{2015ApJ...805..121D} found a strong evidence for massive ($M > 10^{11} M_{\sun}$), quiescent galaxies showing an increasingly important difference between high- and low-density environments at $z \lesssim 1.5$. Also using photometric redshifts \citep[the UKIRT Infrared Deep Sky Survey-Ultra Deep Survey, UKIDSS-UDS, and the Cosmic Assembly Near-infrared Deep Extragalactic Legacy Survey, CANDELS, see][]{2013ApJS..206...10G,2013ApJS..207...24G}, \cite{2015MNRAS.447....2M} found that the GSMF is different in high- and low-density environments up to $z \sim 1.5$.

Although photometric redshifts allow to study the galaxy population at higher redshifts and on larger areas than spectroscopic redshifts (which are usually available on large areas up to $z \sim 1 - 1.5$, or at higher redshifts but on much smaller sky fields and are generally characterized by a low sampling rate), their use is limited by an uncertainty much larger than that of spectroscopic redshifts. Several works have investigated the effect of photometric redshifts on the measurement of the environment \citep[see \textit{e.g.}][]{2016ApJ...825...40L,2016MNRAS.462.1786C,2015MNRAS.451..660E,2015MNRAS.446.2582F,2005ApJ...634..833C,2012MNRAS.419.2670M}. We base this study on our previous work \citep{2016A&A...585A.116M}, in which we extensively tested our methods on mock galaxy catalogues to investigate whether it is still possible to study the GSMF in extreme environments if the density field is measured with photometric redshifts.

In this work we exploit the large statistical sample of the UltraVISTA Survey \citep{2012A&A...544A.156M}, a deep photometric survey performed in the near-infrared in the COSMOS field. In particular, we make use of the high-precision photometric redshift sample of \citet{2013A&A...556A..55I}, which allows us to reach high redshifts with a large enough statistical sample to study the GSMF in different environments for both quiescent and star-forming galaxies. This large statistical dataset allows us to obtain a unique picture of the appearance of environmental signatures in the galaxy mass distribution. In fact, we are able to measure the GSMF in a self-consistent manner, exploring the redshift range $0.2 < z < 3$, analysing different environments and both quiescent and star-forming galaxy populations with a sufficiently faint $K$-band limiting magnitude to reach a rather low mass completeness limit ($M \sim 10^{10} M_{\sun}$). Moreover, the high precision of the photometric redshifts we used has allowed us to obtain a robust measurement of the local density field, clearly distinguishing the most extreme environments (\textit{i.e.} high- and low-density regions) out to high-redshift. This, coupled with the preparatory work described in \citet{2016A&A...585A.116M}, allows us to achieve robust results and to confidently track environmental effects on the GSMF over a large redshift range.

We briefly describe the dataset that we used in Sect. \ref{dataandmethod}. We present our main results in Sect. \ref{results}. We compare our findings with previous works in the literature in Sect. \ref{comparison} and we discuss our results in Sect. \ref{discussion}. We summarize our conclusions in Sect. \ref{conclusions}. A standard cosmology with $\varOmega_{\varLambda} = 0.7$, $\varOmega_{m} = 0.3$ and $H_0 = 70$ km s$^{-1}$ Mpc$^{-1}$ is adopted throughout, together with a \citet{2003PASP..115..763C} IMF.

\section{Data and method}
\label{dataandmethod}
We will briefly review the data set that we used to perform the analysis, together with the methods used to estimate the environment and to calculate the GSMF for the various samples.

\subsection{Sample}
\label{data}
The sample that we used is composed of galaxies from the UltraVISTA Survey \citep{2012A&A...544A.156M}, with photometric redshifts and physical parameters (stellar masses, absolute magnitudes, and restframe colors) derived by \citet{2013A&A...556A..55I}. In particular, photometric redshifts and stellar masses have been measured by fitting to the multi-band photometry synthetic spectra generated using stellar population models \citep{2003MNRAS.344.1000B} and galaxy templates \citep{2007ApJ...663...81P} with the \textit{Le Phare} code \citep{2002MNRAS.329..355A, 2006A&A...457..841I}. A \citet{2000ApJ...533..682C} extinction law has been assumed, while emission line contributions have been modeled after \citet{2009ApJ...690.1236I}. \citet{2013A&A...556A..55I} assumed three metallicity values ($Z = 0.004, 0.008, 0.02$) and exponentially declining star-formation histories in the form of $\tau^{-1}e^{-t/\tau}$ (with $\tau$ values in the range 0.1-30 Gyr). Moreover, \citet{2009ApJ...690.1236I} imposed a low extinction prior on galaxies with low SFR (in the form of $E(B-V) < 0.15$ if age/$\tau > 4$).

The total sample has been selected in $K_{S}$ band and is composed of 339\,384 objects. After the removal of X-ray sources, stars and objects in masked areas, we are left with 209\,758 galaxies with photometric redshift between $0.2 \le z \le 4$, $K_S \le 24$ and measured stellar mass. These objects constitute the final sample on which we performed our analysis. The $K_S$-band and redshift cuts have been performed to be consistent with \citet{2013A&A...556A..55I} and to be able to compare the GSMF for the total, quiescent and star-forming populations with what derived by \citet{2013A&A...556A..55I}. We divided the galaxies of the final sample in quiescent and star-forming following the color-color diagram ($NUV-r^+$ \textit{vs} $r^+-J$) as in \citet{2013A&A...556A..55I}. Compared to a selection based on a $UVJ$ diagram, \citet{2010ApJ...709..644I,2013A&A...556A..55I} argued that the $NUV-r^{+}-J$ color-color diagram allows us to obtain a better distinction between star-forming and quiescent galaxies, as the $NUV-r^{+}$ color is a better indicator of the current (compared to past) star-formation activity \citep[see \textit{e.g.}][]{2007ApJS..173..342M,2007A&A...476..137A}. Moreover, the $NUV$ rest-frame band is still sampled by optical data at $z > 2$ which does not happen for the rest-frame $U$ band. According to the $NUV-r^+$ \textit{vs} $r^+-J$ selection, $\sim 10$\% of the galaxies between $0.2 \le z \le 4$ are quiescent and the remaining fraction of $\sim 90$\% are star-forming. 

We performed our analysis in 8 redshift bins from $z = 0.2$ to $z = 4$. Following \citet{2013A&A...556A..55I} we assumed the photometric redshift uncertainty to be $\sigma_{\varDelta z/(1+z)} = 0.01$ for all the galaxies in the sample. We chose a value of $\sigma_{\varDelta z/(1+z)} = 0.01$ to be consistent with Fig. 1 of \citet{2013A&A...556A..55I}, which shows a comparison between the photometric redshifts for the UltraVISTA Survey and spectroscopic redshifts from a set of various samples up to $K_S \le 24$. Moreover, a value of $\sigma_{\varDelta z/(1+z)} = 0.01$ is in agreement with the average of the error reported in Table 1 of \citet{2013A&A...556A..55I}, weighted by the number of galaxies in each spectroscopic sample used to determine the error. These samples are rather small, sometimes only tens of galaxies, and may therefore overestimate the photometric redshift uncertainty if used independently. Nevertheless, we know that the value we assumed may underestimate the photometric redshift uncertainty at $z \ge 1.5$ and for faint galaxies. For this reason, we have performed a test using a larger photometric redshift error for galaxies at $z > 1.5$. As briefly discussed in Appendix \ref{errorigrandiexplained} we found that our results are not significantly affected by larger photometric redshift uncertainties.

\subsection{Method for the estimation of the environment and the Galaxy Stellar Mass Function}
\label{methodenvironment}
The environment has been determined using a fixed aperture method \citep[similar to what done in][]{2009ApJ...690.1883G}. The performance of this method with photometric redshifts has been extensively tested using mock galaxy catalogues in \citet{2016A&A...585A.116M}. Following what we have found in our previous work, we used a cylinder, centered on each galaxy, with radius $R = 0.3, 0.6, 1,$ and $2$ Mpc and with height $h$ equal to the $3\sigma$ photometric redshift error.
\begin{equation}
h = \pm 1.5 \cdot \sigma_{\varDelta z/(1+z)} \cdot (1+z)
\end{equation}
with $\sigma_{\varDelta z/(1+z)} = 0.01$.

All the galaxies in the sample were used as both targets and tracers for the density field estimation. The measurement of the environment around each galaxy (hereafter defined as target galaxy) was performed by counting how many other galaxies were present inside the cylinder (in the following referred to as tracer galaxies) and then dividing by the cylinder volume. We decided to use volume densities instead of surface densities because they allow us to take the variations of the cylinder volume (due to the variation of the volume height inside the same redshift bin) into account on a galaxy by galaxy basis. In fact, as we chose a cylinder length in the radial direction proportional to the photometric redshift error of each galaxy, galaxies at different redshifts, even inside the same redshift bin, will have different volume sizes. This can create differences in their environment if not properly accounted for. By using volume densities the problem is solved in a self consistent fashion and environmental densities can be better compared \citep[see][for details]{2016A&A...585A.116M}.

The UltraVISTA-COSMOS field has a complicated shape, due to many holes left in the field by saturated stars. Galaxies close to edges or holes in the field can have their environmental measurement biased. In order to limit this effect we applied a correction to the measured environments for galaxies too close to the edges. We rejected all galaxies for which the fraction of the area outside the survey edges (including holes in the field) was greater than 50\% and we corrected the measurement of the density field for all other galaxies by dividing for the fraction of the aperture area inside the edges. Moreover, galaxies with ${\rm R.A.}\:(\deg) > 150.55$ and ${\rm dec}\:(\deg) < 1.8$ were not used in the measurement of the environment, as they lie in a small sky area far from the main field and they would have been too dominated by edge effects. The sample sizes after the correction for the edge effects are reduced to 208\,624, 208\,446, 208\,138, and 207\,183 in the $R = 0.3, 0.6, 1,$ and $2$ Mpc case, respectively. Mass completeness limits for these samples have been calculated as in \citet{2010A&A...523A..13P} and are in very good agreement with those of \citet{2013A&A...556A..55I}. In particular, we select the 20\% faintest galaxies in each redshift bin, separately in the case of the total, quiescent and star-forming galaxy populations. For these galaxies we measure $M_{min}$, the mass that they would have if their apparent magnitude ($m_{K}$) were equal to the limiting magnitude of the UltraVISTA survey in the $K_{S}$ band ($K_{S} = 24$), through the formula $\log (M_{min}) = \log (M) + 0.4(m_{K}-24)$. The mass completeness limit $M_{lim}$ is then the value of $M_{min}$ corresponding to the 90th percentile of the $M_{min}$ distribution. As an example, the mass completeness limits derived with this procedure for the total, passive and star-forming populations range from $(M_{lim,tot},M_{lim,q},M_{lim,sf}) = (10^{8.5}, 10^{9}, 10^{8.5})$ at $z \sim 0.5$ to $(10^{9.2}, 10^{9.5}, 10^{9.2})$ at $z \sim 1$ and $(10^{9.7}, 10^{10.2}, 10^{9.7})$ at $z \sim 2$.

High-density and low-density environments were selected as those above the 75th percentile or below the 25th percentile of the volume density distribution of galaxies with $M^{\ast} \ge 10^{10} M_{\sun}$, with the quartiles of the distribution computed at each redshift bin. We chose this mass threshold because the increase in the mass completeness limit of our sample with redshift can influence the density value of the percentiles used to define high- and low-density environments. In fact, at low redshifts our sample is complete at lower masses, the dynamic range of the density measurement is large and the environmental density thresholds used to define high- and low-density environments are lower. Conversely at high redshifts the dynamic range is reduced, because the sample is complete only at higher masses compared to the low redshift case, and the threshold for the definition of high- and low-density environments is higher. The volume density distribution is sensitive to the mass completeness limit of the sample, because of the mass-density relation. In this way we would not be able to compare the same kind of environments at low and high redshifts. By chosing a mass cut close to the mass completeness limit of the highest redshift bin we are then able to compare the galaxy population at low and high redshift in a consistent way. High- and low-density environments have been defined for both quiescent and star-forming galaxies using the quartiles of the total galaxy population. In the following, we will use the notation $D_{75}$ and $D_{25}$ to refer to high- and low-density environments, respectively.

Galaxy Stellar Mass Functions have been calculated with the non parametric $1/V_{max}$ estimator \citep{1980ApJ...235..694A}. They have been calculated separately for quiescent and star-forming galaxies, both in high-density and low-density environments. A comparison of our GSMF and those of \citet{2013A&A...556A..55I} shows perfect agreement. As the thresholds for defining high- and low-density environments have been calculated using only galaxies more massive than $10^{10} M_{\sun}$, the GSMF result roughly normalized at high masses, due to the fact that

\begin{equation}
\int_{M \ge 10^{10} M_{\sun}} \varPhi_{D75}(M)dM = \int_{M \ge 10^{10} M_{\sun}} \varPhi_{D25}(M)dM
\end{equation}

When calculating mass functions, if the number of galaxies in a given mass bin is lower than two, we applied the prescription for small counts Poisson statistics of \citet{1986ApJ...303..336G}, in the form of $1\sigma$ upper and lower limits of Tables 1 and 2. In particular, these tables present the number of true Poisson events corresponding to a rate of one or zero observed events. In the case of one observed event (mass bins containing only one galaxy), the number of real Poisson events can be used to correct the error measurement. In case of empty mass bins (zero observed events), the number of real Poisson events divided by the volume corresponding to the selected redshift bin can be used to set an upper limit for the GSMF in a larger mass range.

\section{Results}
\label{results}
Although the analysis presented in this work is based on photometric redshifts, the method that we used to reconstruct local density and that we tested on mock galaxy catalogues in our previous work \citep{2016A&A...585A.116M} is able to provide us with a robust measurement of galaxy environments up to $z = 3$ and on various physical scales. The high-precision photometric redshifts of the UltraVISTA sample allow us to trace environmental effects on galaxy properties throughout cosmic history, contributing in the creation of a consistent picture of galaxy evolution. In the following, only the cases with $R = 0.3$ and $R = 2$ Mpc will be discussed at length. The other values of $R$, which constitute intermediate cases between those reported here, have been analyzed, but will not be reported for the sake of clarity and conciseness.

\begin{figure}
\centering
\includegraphics[width = \linewidth]{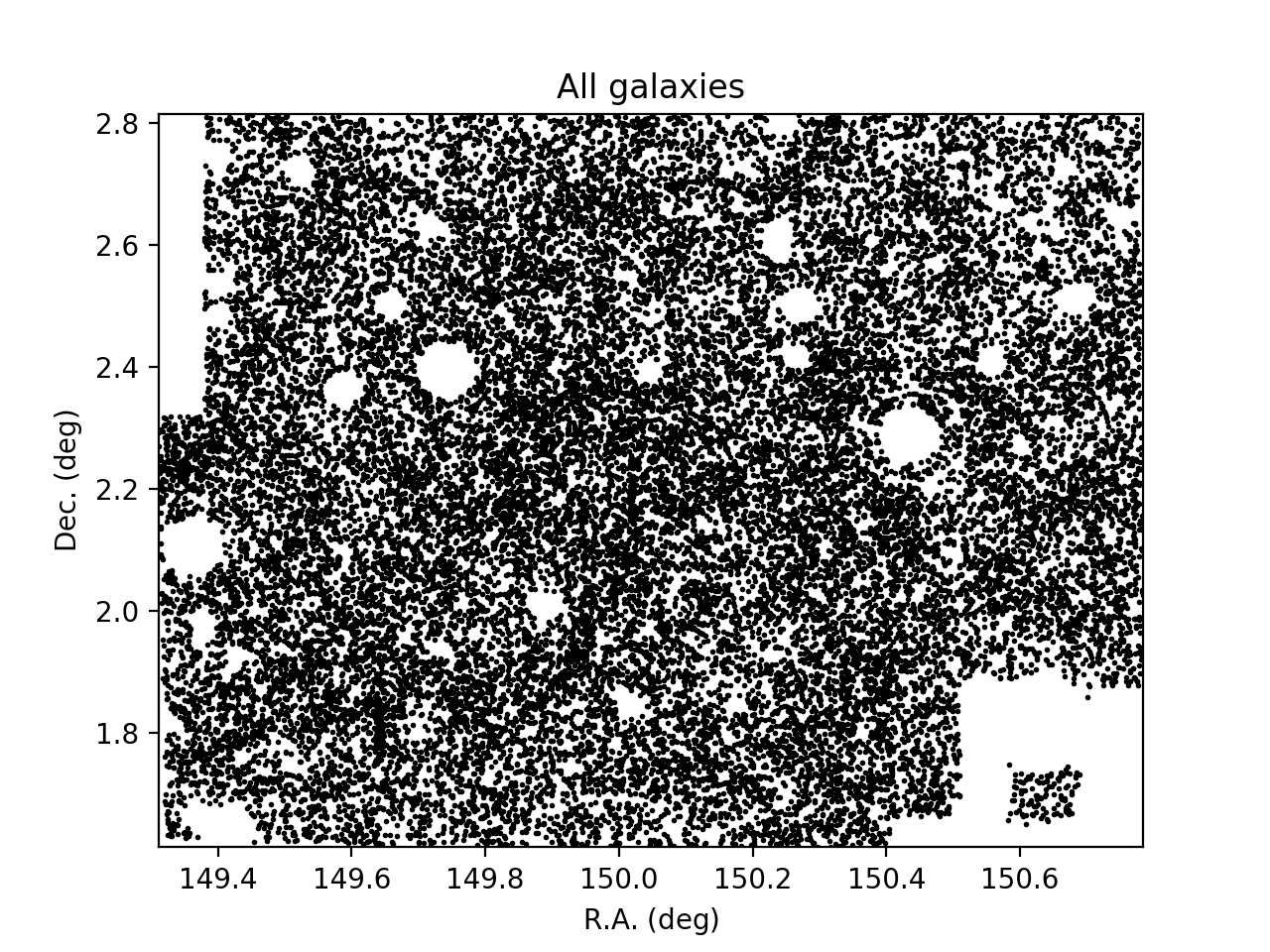}
\includegraphics[width = \linewidth]{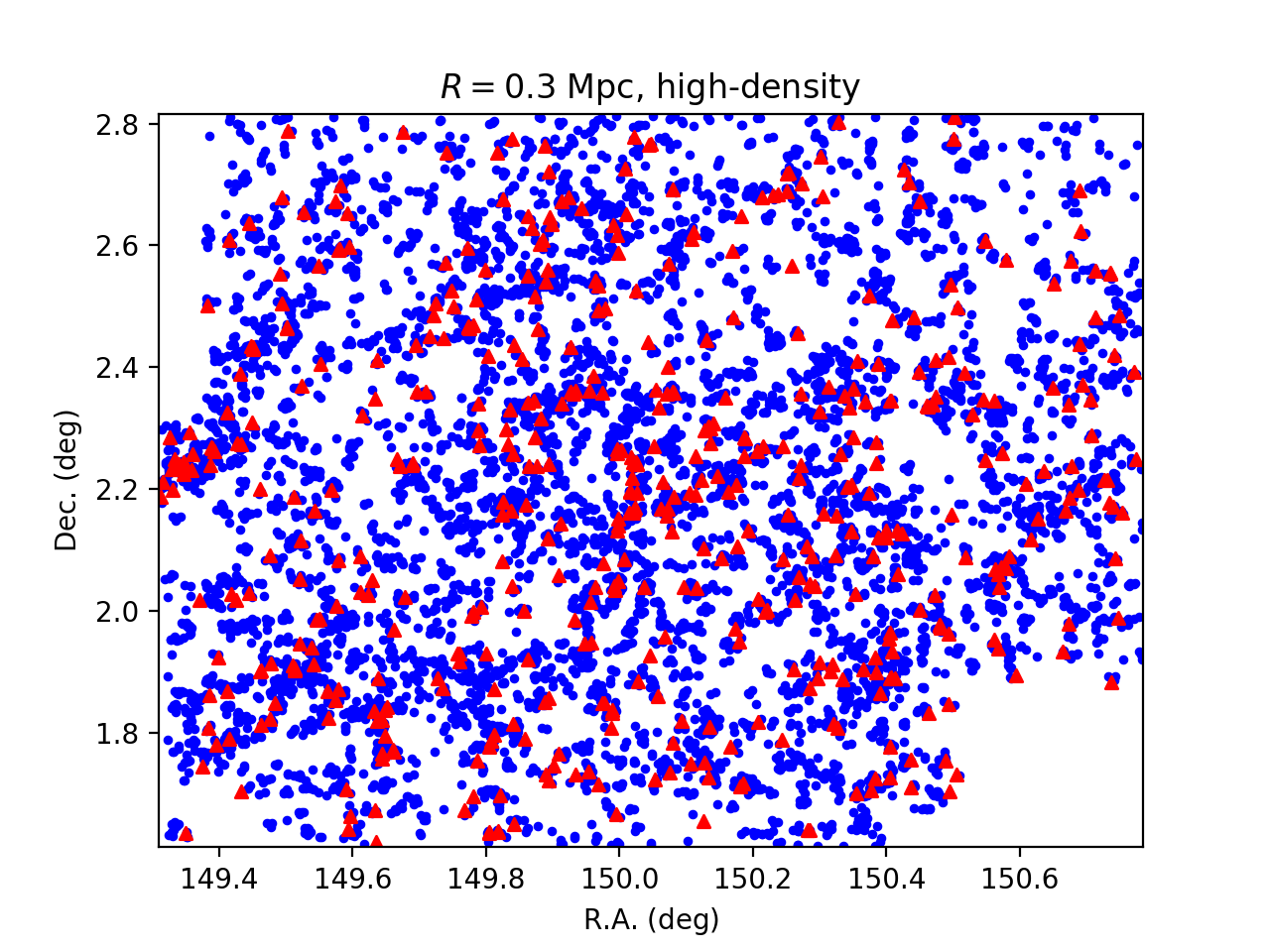}
\includegraphics[width = \linewidth]{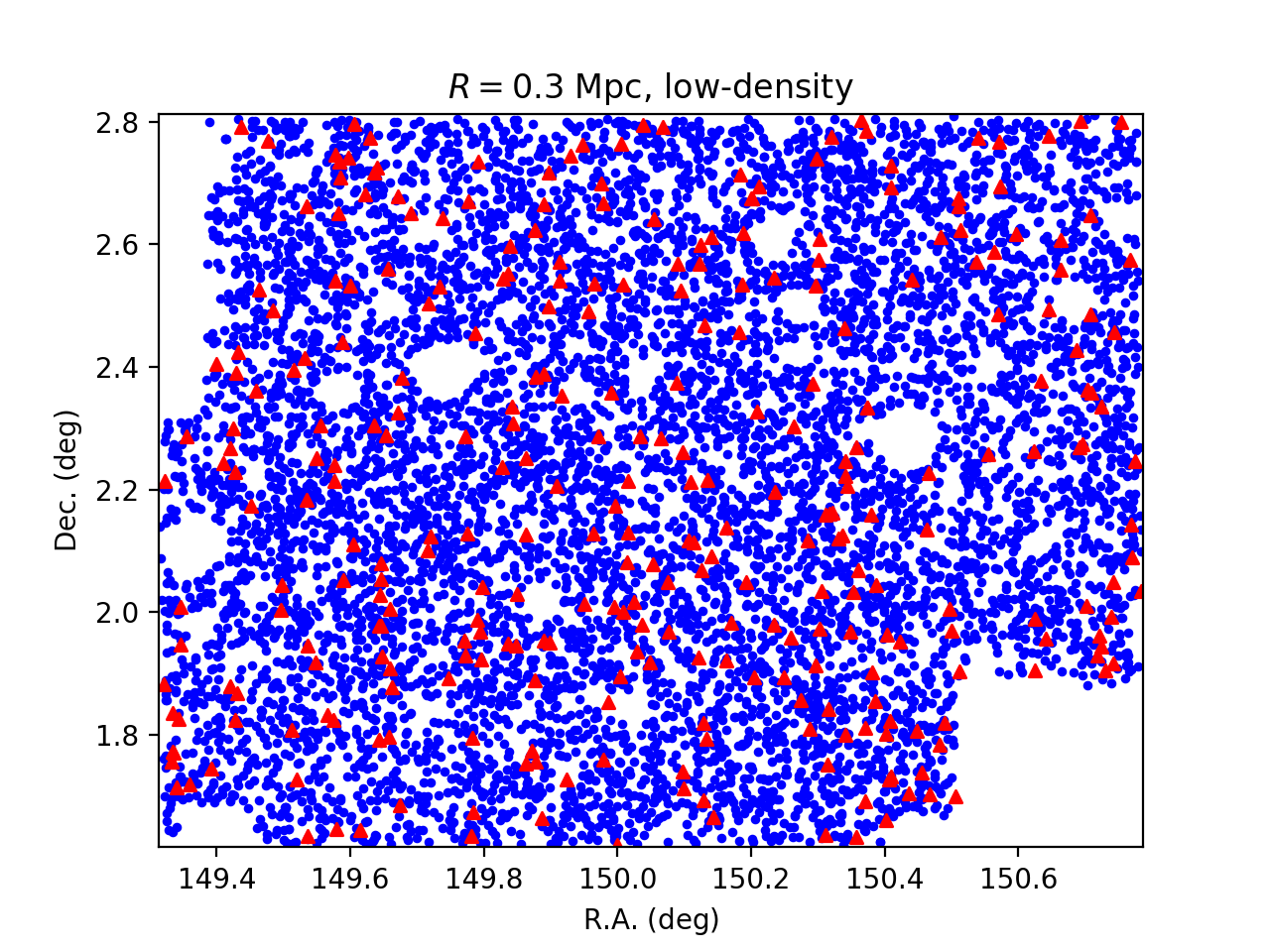}
\caption{\textit{Sky maps of the UltraVISTA field}. Only the case for $R = 0.3$ Mpc, $z \in [1.5,2.0]$ is shown. Top panel: black dots represent the total UltraVISTA sample in the considered redshift bin. Middle and bottom panel are only high- and low-density environments, respectively. Red triangles and blue dots represent quiescent and star-forming galaxies. The fixed aperture method is able to identify galaxies in different environments at high redshift. Quiescent galaxies are located preferentially in high-density environments.}
\label{clusters1}
\end{figure}

Figure \ref{clusters1} shows an example of the performance of the fixed aperture method in estimating the density field. This figure shows the UltraVISTA sky field in a representative high-redshift bin ($1.5 \le z \le 2.0$) for the total UltraVISTA sample and only high- and low density environments separately. Red and blue dots refer to quiescent and star-forming galaxies, respectively. Only the fixed aperture radius $R = 0.3$ Mpc is represented, as an example. It can be seen that the fixed aperture method that we implemented is able to identify galaxies in different environments. Galaxies belonging to high-density environments tend to be more clustered, while low-density galaxies appear spatially distributed in a more uniform fashion.

\begin{figure*}
\centering
\includegraphics[width=17cm]{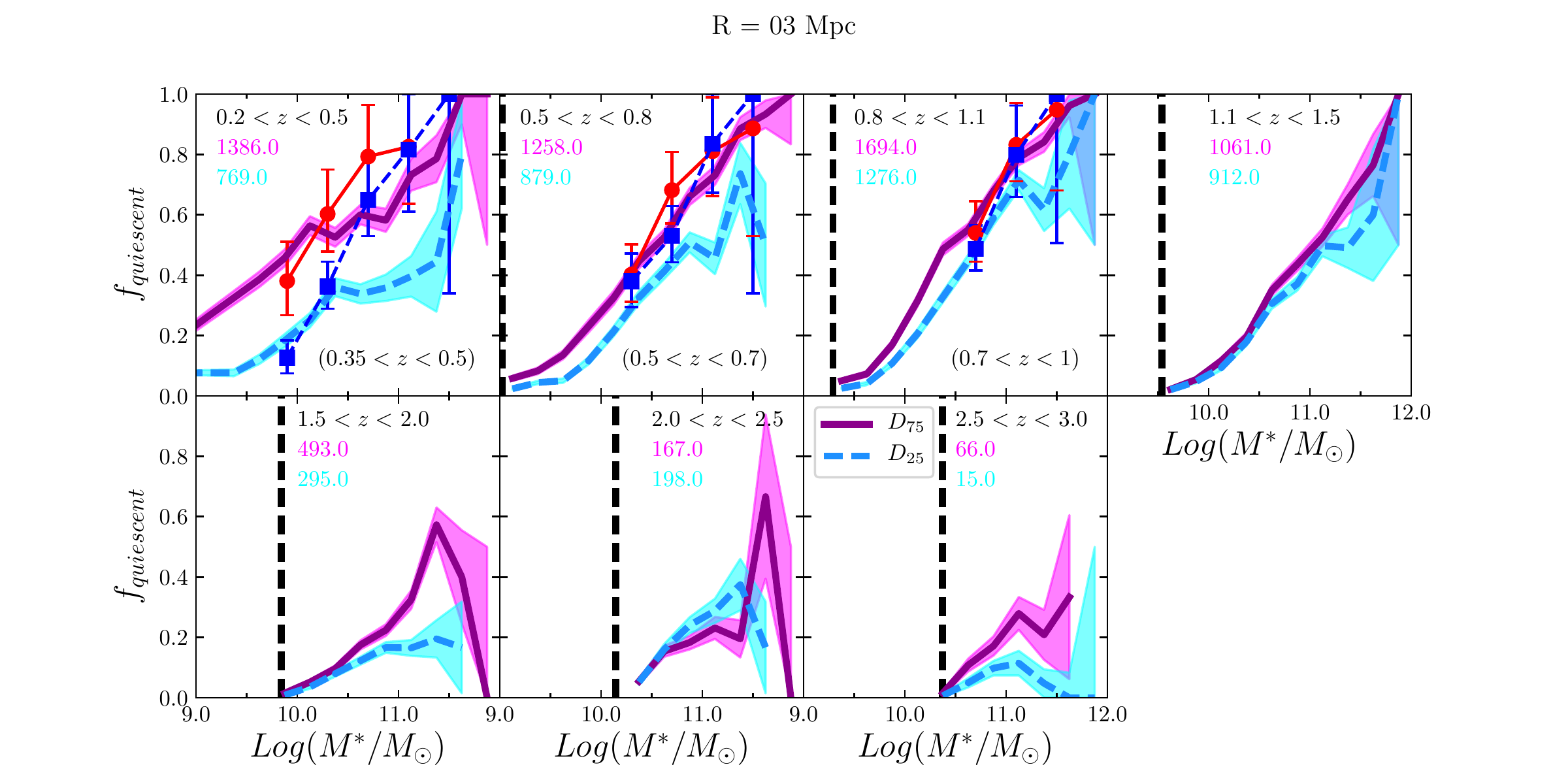}
\includegraphics[width=17cm]{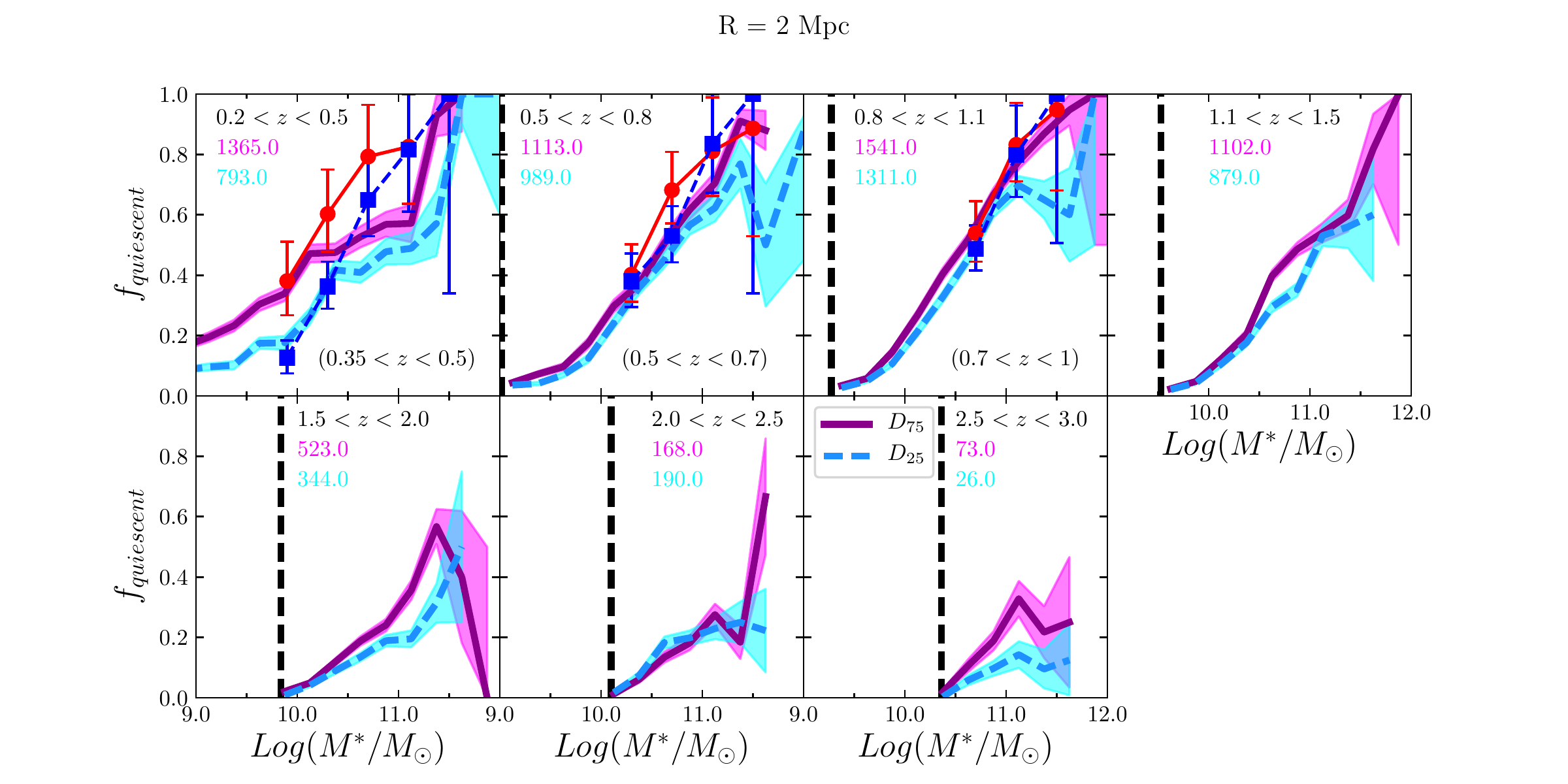}
\caption{\textit{Fraction of quiescent galaxies}. The solid magenta line refers to high-density environments, the dashed cyan line to low-density environments. The shaded regions correspond to the propagated errors on the fraction. For each redshift bin, the number of galaxies above the mass limit in the two environments (magenta for high-density environments and cyan for low-density environments) are reported. The vertical black dashed line corresponds to the mass completeness limit. Top panel refers to a fixed aperture radius of $R = 0.3$ Mpc, bottom panel to $R = 2$ Mpc. In the first three redshift bins, the fractions of quiescent galaxies in high- and low-density environments from the work by \citet[][see their figure 8]{2010A&A...524A..76B} are reported for comparison above the mass completeness limit. Red circles and solid lines refer to high-density environments, blue squares and dashed lines to low-density environments. The redshift bins in which the fractions of \citet{2010A&A...524A..76B} have been calculated are reported in parentheses in the bottom right corners of the plots. The fraction of quiescent galaxies is larger in high-density environments up to $z \sim 2$.}
\label{ETGFraction}
\end{figure*}

Interestingly, it can be seen how, although rare at this redshift, quiescent galaxies tend to be slightly more visible in the high-density regions compared to the low-density ones. This trend can be expressed quantitatively by looking at the fraction of quiescent galaxies as a function of environment, redshift and mass (shown in Figure \ref{ETGFraction}). As expected the fraction of quiescent galaxies increases with cosmic time in both environments. Nevertheless, these fractions show how quiescent galaxies are more numerous in high-density environments compared to low-density ones as a function of mass. Although the difference is a function of mass and redshift, it remains well visible up to $z \sim 2$ for both the $R = 0.3$ Mpc and the $R = 2$ Mpc case. For masses $\sim 10^{11} M_{\sun}$, at $z \sim 0.5$ 60\% of the galaxies in high-density environments are quiescent, while only 40\% in low-density environments. At $z \sim 1$ the difference is reduced to $\lesssim 10$\%, but it is still visible. In the $R = 2$ Mpc case, differences of $\sim 10$\% at $z \sim 0.5$ are reduced to $\sim 5$\% at $z \sim 1$. 

The trend visible in our data is in agreement also with what found in other works, using both global environment \citep[see \textit{e.g.}][who performed analysis using various samples of clusters at $z \sim 1-1.5$]{2013A&A...557A..15V,2016A&A...592A.161N,2012ApJ...746..188M} and local environment definitions. In particular, with respect to local environment, \citet{2006MNRAS.373..469B} found a fraction of quiescent galaxies $\sim 20\%$ higher in high-density environments for masses of  $\sim 10^{11} M_{\sun}$ at $z \lesssim 0.1$, compared to the lowest densities they explored. Results in agreement with our fractions of quiescent galaxies are found also by \citet{2016ApJ...825..113D}, who reported a fraction of quiescent galaxies higher by $\sim 20-40\%$ for masses of $\sim 10^{11} M_{\sun}$ at $z \sim 0.5$ and by $\lesssim 20\%$ at $z \sim 1$ in high density environments compared to low-density ones in the COSMOS field \citep[see also][]{2013ApJS..206....3S,2016ApJ...817...97L}. For direct comparison, in Figure \ref{ETGFraction} also the fractions of quiescent galaxies in different environments for the zCOSMOS Survey \citep[][see their figure 8]{2010A&A...524A..76B} are reported (for a more detailed comparison between our work and \citealt{2010A&A...524A..76B} see section \ref{comparison}). 

\subsection{The GSMF of the UltraVISTA sample}
\begin{figure*}
\centering
\includegraphics[width = 17cm, trim = 0cm 0.5cm 0cm 0.5cm, clip]{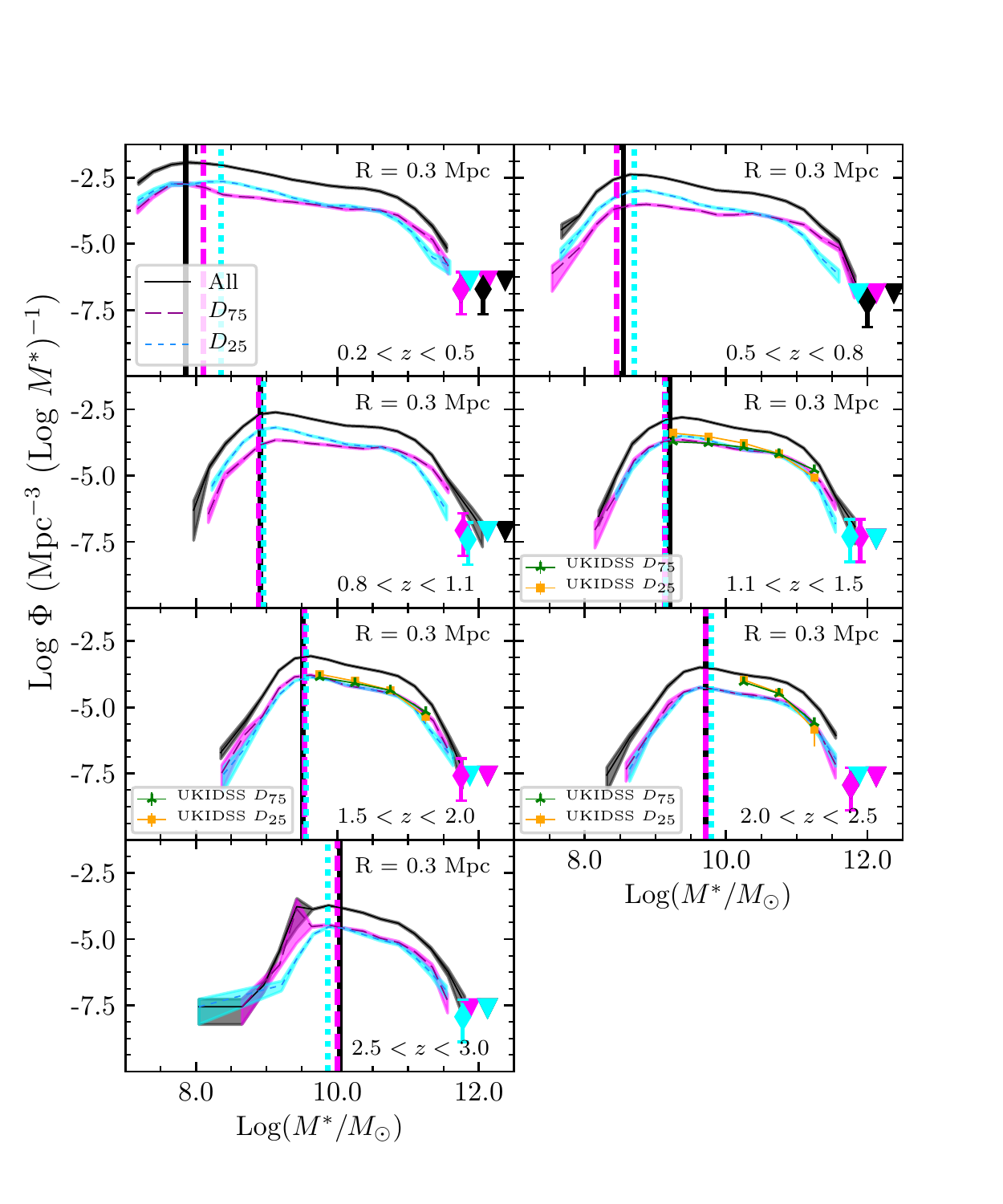}
\caption{\textit{GSMF of UltraVISTA galaxies - all galaxies}. The solid black curve refers to the total GSMF for all UltraVISTA galaxies, the long-dashed magenta curve refers to high-density environments and the short-dashed cyan curve refers to low-density environments. Vertical lines are the mass completeness limits, style- and colour-coded as the corresponding GSMF. Diamonds represent mass bins with only one galaxy, downward triangles are upper limits for mass bins with zero galaxies. Shaded areas represent Poissonian errors. In the bins at redshift $1.1 < z < 1.5$, $1.5 < z < 2.0$, and $2.0 < z < 2.5$ green stars and orange squares correspond to the UKIDSS-CANDELS GSMF \citep[][see their figure 8, green stars refer to high-density environments, orange squares to low-density environments]{2015MNRAS.447....2M}. For the \citet{2015MNRAS.447....2M} GSMF, only points above their mass completeness limit have been considered. The redshift bins in which the \citet{2015MNRAS.447....2M} GSMFs have been calculated are $1.0 < z < 1.5$, $1.5 < z < 2.0$, and $2.0 < z < 2.5$. The high-mass end of the UltraVISTA GSMF is enhanced in high-density environments, while the low-mass end is depleted with respect to low-density environments.}
\label{MFall}
\end{figure*}

Figure \ref{MFall} shows a first example of the GSMF for all UltraVISTA galaxies. In the same Figure also the GSMF for high-density and low-density environments are shown, in the case of $R = 0.3$ Mpc. We do not report the GSMF for other radii for the sake of clarity, although when performing the analysis on the shape of the GSMF in different environments all apertures will be considered for completeness. It can be seen how the GSMF of high-density and low-density environments are different. The high-mass end of the GSMF (above $M = 10^{10.5 - 11} M_{\sun}$) is enhanced in the case of high-density environments, in comparison to low-density ones, while the low mass end is depleted. This difference can be appreciated up to $z \sim 2$, where no more differences can be seen between high-density and low-density environments.

\begin{figure*}
\centering
\includegraphics[width = 17cm, trim = 0cm 0.5cm 0cm 0.5cm, clip]{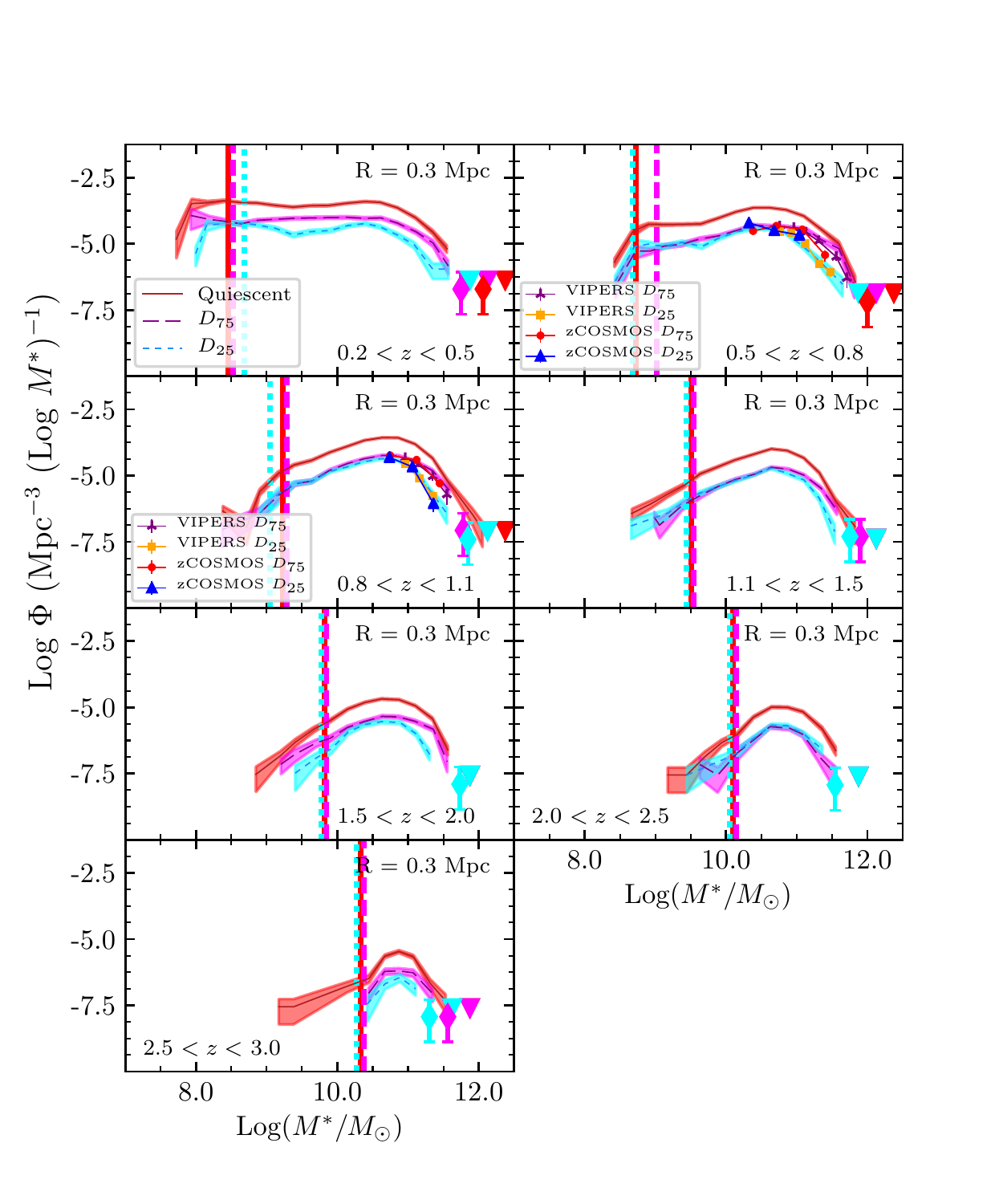}
\caption{\textit{GSMF of UltraVISTA galaxies - quiescent galaxies}. The solid red curve refers to the total GSMF for quiescent galaxies only, the long-dashed magenta curve refers to high-density environments and the short-dashed cyan curve refers to low-density environments. Vertical lines are the mass completeness limits, style- and colour-coded as the corresponding GSMF. Diamonds represent mass bins with only one galaxy, downward triangles are upper limits for mass bins with zero galaxies. Shaded areas represent Poissonian errors. In the bins at redshift $0.5 < z < 0.8$ and $0.8 < z < 1.1$ red circles and blue upward triangles correspond to the zCOSMOS GSMF \citep[][see their figure 5, red circles represent high-density environments and blue upward triangles represent low density environments]{2010A&A...524A..76B}, for quiescent galaxies. Purple stars and orange squares correspond to the VIPERS GSMF \citep[][see their figure 4, purple stars represent high-density environments, orange squares represent low-density environments]{2016A&A...586A..23D}, for quiescent galaxies. For zCOSMOS and VIPERS GSMF only points above the respective mass completeness limits are shown. The redshift bins in which zCOSMOS GSMFs have been calculated are $0.5 < z < 0.7$ and $0.7 < z < 1.0$, the redshift bins in which VIPERS GSMFs have been calculated are $0.65 < z < 0.8$ and $0.8 < z < 0.9$. The high-mass end of the UltraVISTA quiescent GSMF is enhanced in high-density environments up to $z \sim 2$, while no difference seems to be present at the intermediate- and low-mass end.}
\label{MFquiescent}
\end{figure*}

\begin{figure*}
\centering
\includegraphics[width = 17cm, trim = 0cm 0.5cm 0cm 0.5cm, clip]{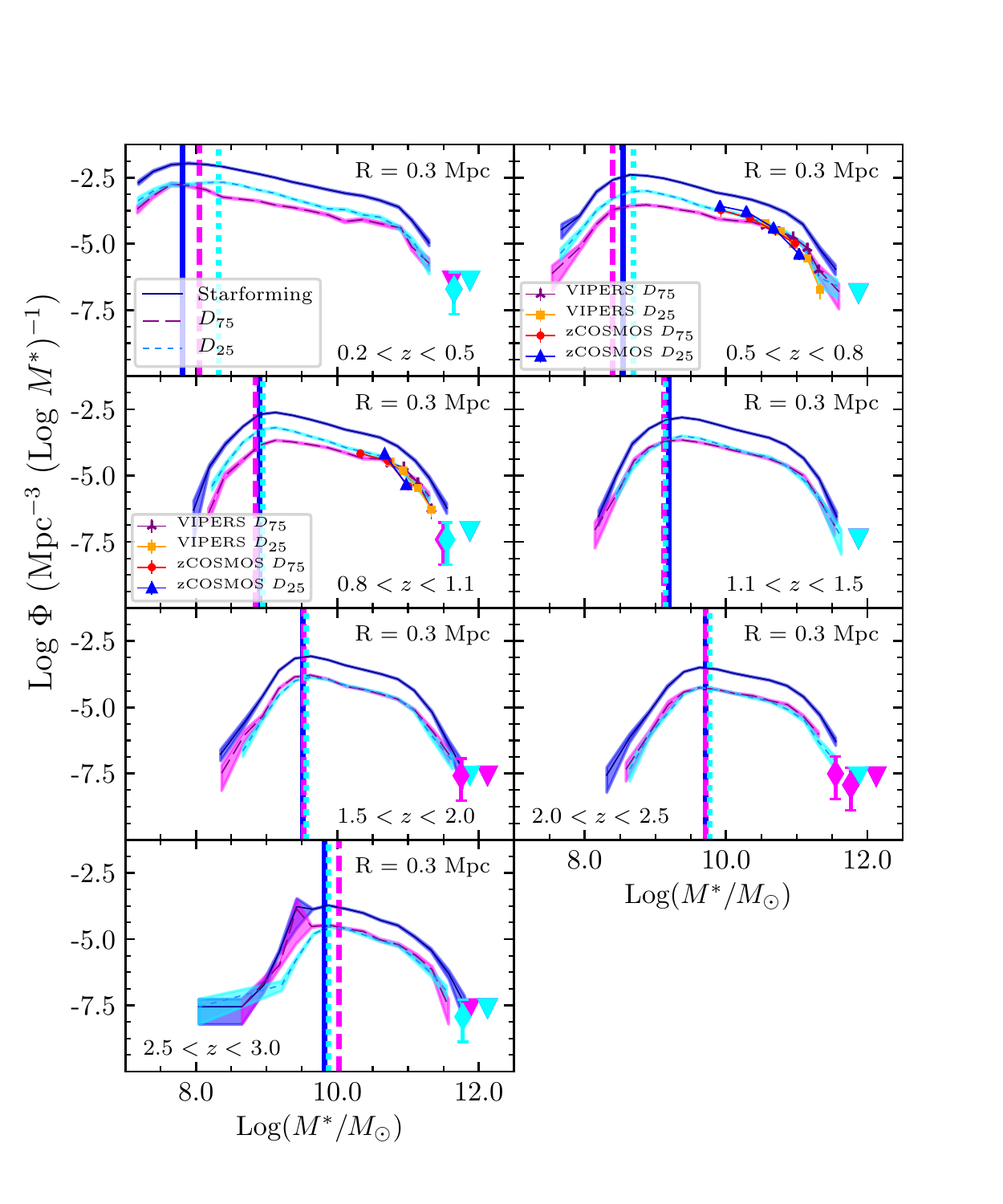}
\caption{\textit{GSMF of UltraVISTA galaxies - star-forming galaxies}. The solid blue curve refers to the total GSMF for star-forming galaxies only, the long-dashed magenta curve refers to high-density environments and the short-dashed cyan curve refers to low-density environments. Vertical lines are the mass completeness limits, style- and colour-coded as the corresponding GSMF. Diamonds represent mass bins with only one galaxy, downward triangles are upper limits for mass bins with zero galaxies. Shaded areas represent Poissonian errors. In the bins at redshift $0.5 < z < 0.8$ and $0.8 < z < 1.1$ red circles and blue upward triangles correspond to the zCOSMOS GSMF \citep[][see their figure 5, red circles represent high-density environments and blue upward triangles represent low density environments]{2010A&A...524A..76B}, for star-forming galaxies. Purple stars and orange squares correspond to the VIPERS GSMF \citep[][see their figure 4, purple stars represent high-density environments, orange squares represent low-density environments]{2016A&A...586A..23D}, for star-forming galaxies. For zCOSMOS and VIPERS GSMF only points above the respective mass completeness limits are shown. The redshift bins in which zCOSMOS GSMFs have been calculated are $0.5 < z < 0.7$ and $0.7 < z < 1.0$, the redshift bins in which VIPERS GSMFs have been calculated are $0.65 < z < 0.8$ and $0.8 < z < 0.9$. The UltarVISTA star-forming GSMF shows an excess of low-mass galaxies in low density environments up to $z \sim 1.5$, while no difference seems to be present in the high-mass end.}
\label{MFstarforming}
\end{figure*}

If we divide the galaxy population into quiescent and star-forming galaxies, we see how the difference between high- and low-density environments affects different parts of the GSMF in the case of the quiescent galaxy population (Figure \ref{MFquiescent}) and in the case of the star-forming galaxy population (Figure \ref{MFstarforming}). These figures show the GSMF for the quiescent and star-forming components of the total GSMF in high- and low-density environments. For quiescent galaxies, the enhancement of the high-mass end in high-density environments is visible in comparison to low-density environments up to $z \sim 2$. Moreover, a steep decline at low masses is visible. This decline is visible in the high-density and low-density GSMFs and is found also by \citet{2013A&A...556A..55I, 2013ApJ...777...18M, 2015MNRAS.447....2M}. For the star-forming population, instead, the difference is mainly present at low masses (below $10^{11} M_{\sun}$) and at lower redshifts (below $z \sim 1.5$). In these figures, the GSMFs from \citet{2016A&A...586A..23D,2010A&A...524A..76B,2015MNRAS.447....2M} are reported for reference and a more detailed comparison between these works and our results will be carried out in section \ref{comparison}.

\begin{figure*}
\centering
\includegraphics[width=17cm]{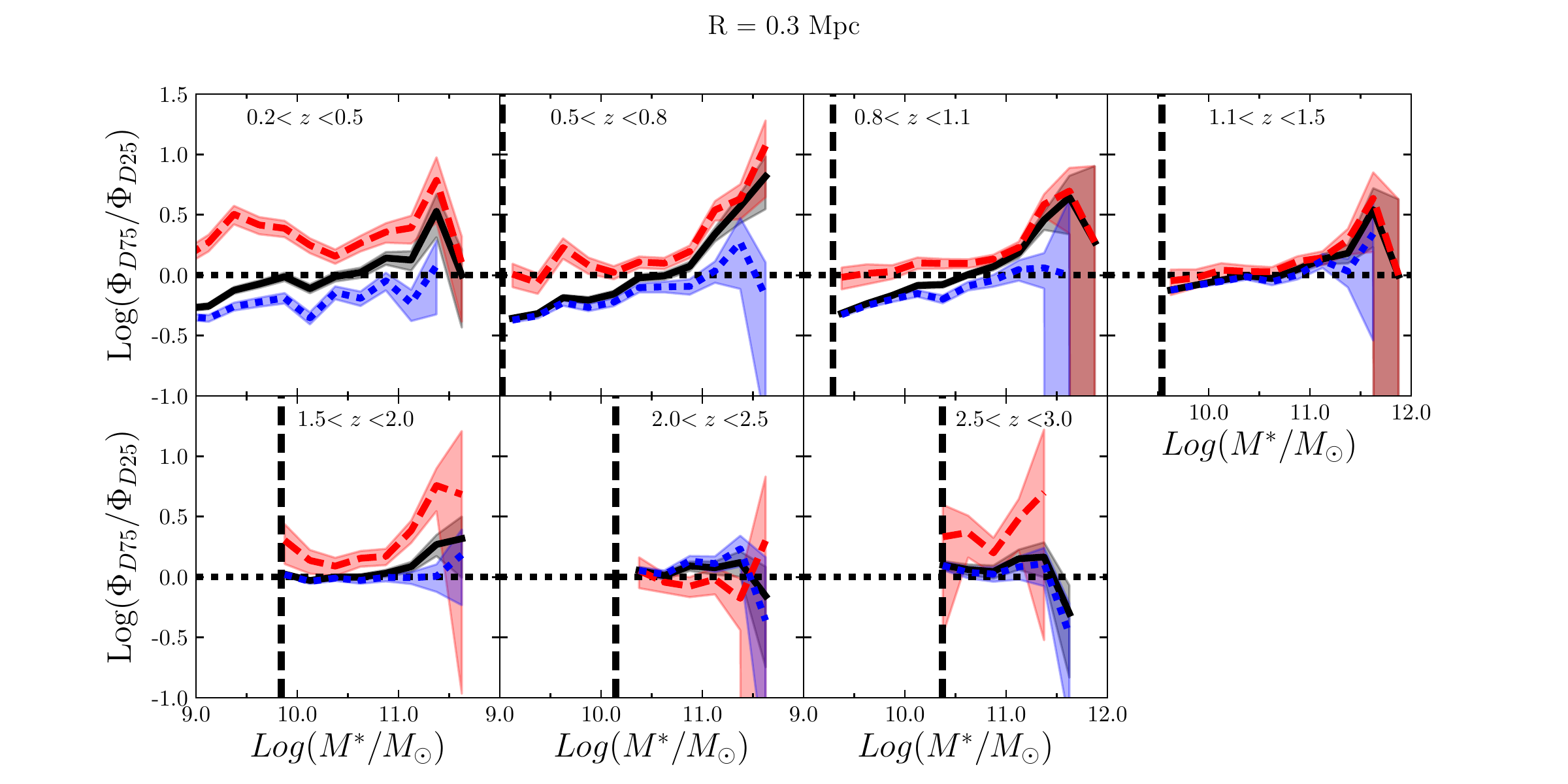}
\includegraphics[width=17cm]{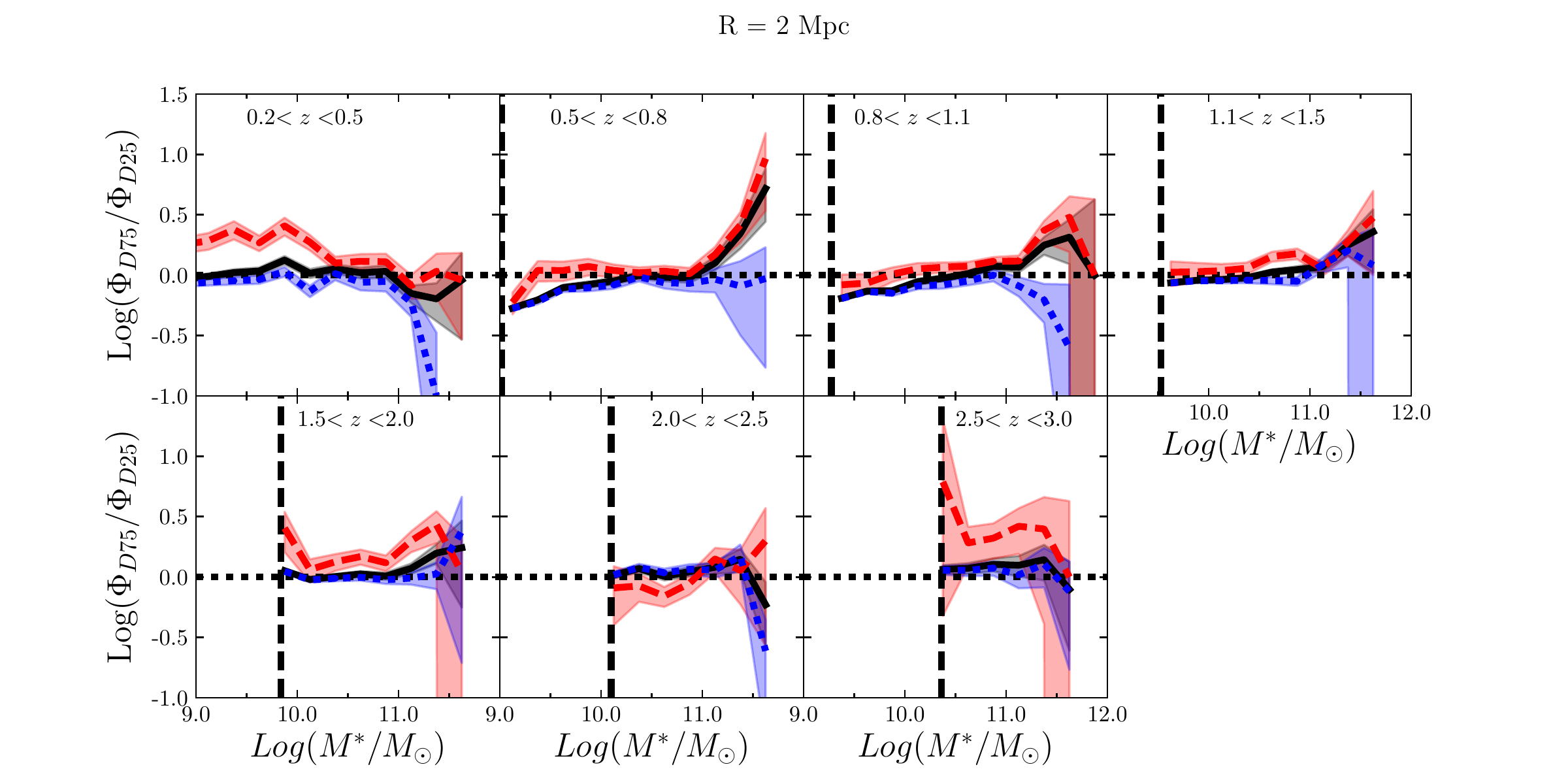}
\caption{\textit{Ratio of high- to low-density GSMF}. Ratio of the high-density ($\varPhi_{D75}$) to the low-density GSMF ($\varPhi_{D25}$) as a function of mass and redshift. The solid black line refers to all galaxies, the long-dashed red line to quiescent galaxies and the short-dashed blue line to star-forming galaxies. The shaded regions correspond to the propagated errors on the ratio. The vertical black dashed line corresponds to the mass completeness limit. Top panel refers to a fixed aperture radius of $R = 0.3$ Mpc, bottom panel to $R = 2$ Mpc. The fact that the logarithm of the ratio of the high-density to the low-density GSMF is generally $\gtrsim 0$ for quiescent galaxies and generally $\lesssim 0$ for star-forming galaxies can be interpreted as quiescent galaxies being more represented in high-density environments and star-forming galaxies being more present in low-density environments.}
\label{Ratios}
\end{figure*}

A more quantitative analysis of the differences between high- and low-density GSMF for the different galaxy populations can be performed by taking the ratios of the high-density to the low-density GSMF for the total, the quiescent and the star-forming galaxy populations as a function of mass and redshift (Figure \ref{Ratios}). In the quiescent and star-forming case the ratios are calculated using the quiescent and star-forming component of the total GSMF in high- and low-density environments. For this reason, the ratio can be greater than 1 (logarithm of the ratio greater than 0, in the figure). It can be seen how the ratio of the high-density to low-density GSMF is typically higher in the case of quiescent galaxies compared to star-forming ones, at least up to $z \sim 2$ for both the $R = 0.3$ Mpc and the $R = 2$ Mpc case. The ratio of high-density to low-density GSMF is generally $\gtrsim 1$ for quiescent galaxies (logarithm of the ratio $\gtrsim 0$) and it is generally $\lesssim 1$ for star-forming galaxies (logarithm of the ratio $\lesssim 0$, in the figure). This can be interpreted as quiescent galaxies being more represented in high-density environments and star-forming galaxies being more present in low-density environments. These ratios also show a trend with mass, both for quiescent and star-forming galaxies. High-density environments are dominated by a more massive galaxy population, and this is generally true for both quiescent and star-forming galaxies. Instead the ratio of high-density to low-density GSMF for the total galaxy population follows the same ratio of star-forming galaxies at low masses and the one of quiescent galaxies at high masses, as expected.

\subsection{The shape of the GSMF in different environments}
\begin{figure*}
\centering
\resizebox{\hsize}{!}{\includegraphics{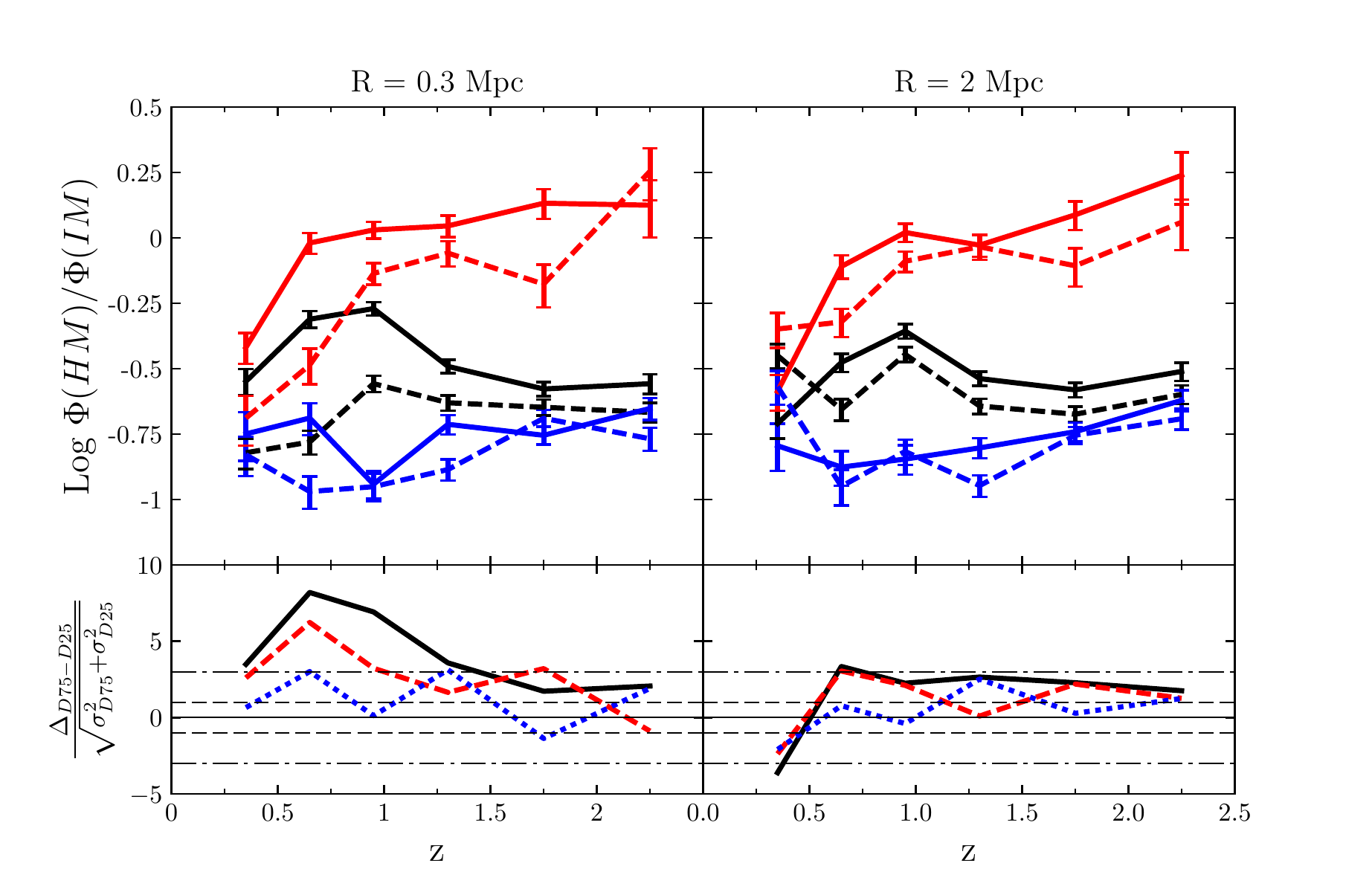}}
\caption{\textit{Shape of the GSMF}. Top panel shows the ratio of the high-mass to the intermediate-mass end of the GSMF (see Eq. \eqref{Mratio}) as a function of redshift. Solid lines refer to high-density environments, dashed lines to low-density ones. Red lines represent quiescent galaxies and blue lines star-forming galaxies. In black we report the ratio for the total galaxy population. Bottom panel shows the difference between the high-density and the low-density case of the curves reported in the top panel, normalized to the sum of the errors (long-dashed red, short-dashed blue, and solid black curves refer to the quiescent, star-forming, and total galaxy populations, respectively). For reference, values corresponding to a difference of $0$, $\pm 1\sigma$, and $\pm 3\sigma$ are reported as thin solid, dashed, and dot-dashed horizontal lines. These ratios quantitatively indicate that the high-mass end of the GSMF is enhanced in high-density environments for quiescent galaxies and not for star-forming ones up to $z \sim 2$.}
\label{massratios}
\end{figure*}

Differences between the shape of the GSMF in high-density and low-density environments can be better seen by taking the ratio of the high-mass end to the intermediate-mass end of the GSMF. In particular, we calculated the quantity
\begin{equation}\label{Mratio}
\log \frac{\Phi(HM)}{\Phi(IM)} = \log \frac{\int_{\log(M) \in [11,11.5]} \varPhi(M)dM}{\int_{\log(M) \in [10,10.5]} \varPhi(M)dM}
\end{equation}
for both quiescent and star-forming galaxies in both high-density and low-density environments (shown in Figure \ref{massratios}). To calculate the ratio we did not include upper limits due to mass bins with zero galaxies, but we did include mass bins with only one count. Moreover, the analysis has been performed only up to $z \sim 2.5$ as the sample of quiescent galaxies begins to be incomplete at $10 \le \log(M/M_{\sun}) \le10.5$ in the last redshift bin.

These ratios show clearly how the difference between high-density and low-density environments is present mainly for quiescent galaxies, rather than for star-forming galaxies. For quiescent galaxies, the ratio of the high-mass to the intermediate-mass end of the GSMF is higher in high-density environments compared to low-density ones. This ratio also shows a trend with redshift, monotonically increasing up to $z \sim 2 $ (for high-density, quiescent galaxies the trend with redshift is more evident for the $R = 2$ Mpc case). This reflects the gradual build-up of the intermediate mass part of the GSMF with cosmic time for the quenched galaxy population, and is in agreement with a scenario in which massive galaxies became passive at earlier times than lower mass galaxies (downsizing). The difference between high-density and low-density environments seems to be present both for small and for large radii, with no significant differences among them. In the bottom panel of the same figure, we report the difference between the high-density and the low-density curves from the top panel, normalized to the sum in quadrature of their errors (in the case where the errors on the ratios are asymmetric, the largest of the two has been considered). These plots show how the difference between high- and low-density environments reaches and largely exceeds $3\sigma$ (being as large as $5\sigma$ in the $R = 0.3$ Mpc case) for quiescent galaxies, while it is lower for star-forming galaxies (mostly of the order of $2\sigma$ and even lower in the $R = 2$ Mpc case).

\begin{figure*}
\centering
\resizebox{\hsize}{!}{\includegraphics{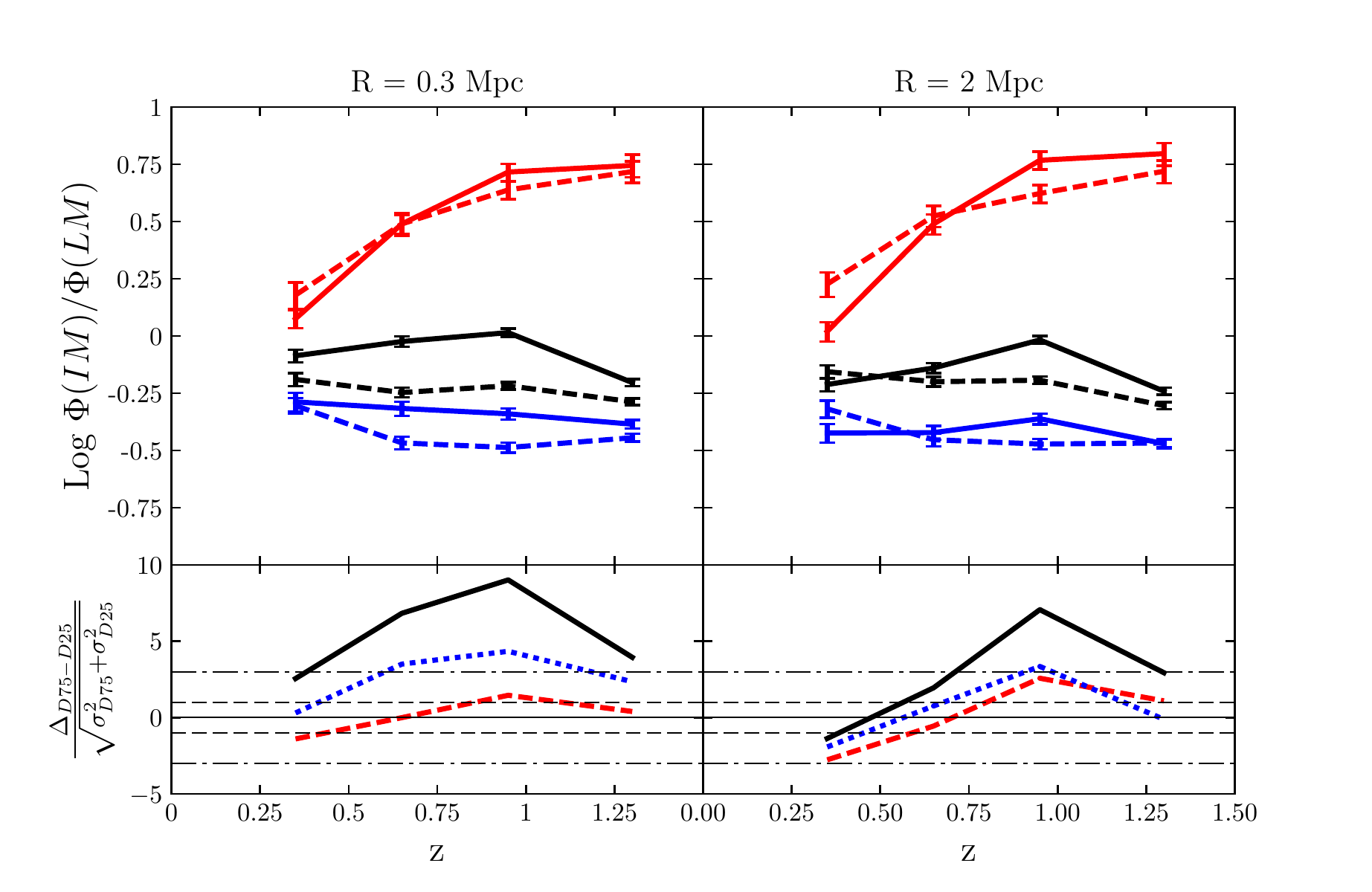}}
\caption{\textit{Shape of the GSMF}. Top panel shows the ratio of the intermediate-mass to the low-mass end of the GSMF (see Eq. \eqref{MratioLMLZ}) as a function of redshift. Solid lines refer to high-density environments, dashed lines to low-density ones. Red lines represent quiescent galaxies and blue lines star-forming galaxies. In black we report the ratio for the total galaxy population. Bottom panel shows the difference between the high-density and the low-density case of the curves reported in the top panel, normalized to the sum of the errors (long-dashed red, short-dashed blue, and solid black curves refer to the quiescent, star-forming, and total galaxy populations, respectively). For reference, values corresponding to a difference of $0$, $\pm 1\sigma$, and $\pm 3\sigma$ are reported as thin solid, dashed, and dot-dashed horizontal lines. These ratios quantitatively indicate that the low-mass end of the star-forming GSMF is depleted in high-density environments up to $z \sim 1.5$, while no difference is present for quiescent galaxies.}
\label{massratioslmlz}
\end{figure*}

Conversely, if we take the ratio of the intermediate-mass end to the low-mass end of the GSMF, differences emerge only for star-forming galaxies at $z \le 1.5$. In particular, we calculated the quantity
\begin{equation}\label{MratioLMLZ}
\log \frac{\Phi(IM)}{\Phi(LM)} = \log \frac{\int_{\log(M) \in [10.5,11]} \varPhi(M)dM}{\int_{\log(M) \in [9.5,10]} \varPhi(M)dM}
\end{equation}
for both quiescent and star-forming galaxies in both high-density and low-density environments (shown in Figure \ref{massratioslmlz}). Again, to calculate the ratio we did not include upper limits due to mass bins with zero galaxies, but we did include mass bins with only one count. The quantity described in Eq \eqref{MratioLMLZ} has been computed only up to $z \sim 1.5$ as for higher redshifts the sample of quiescent galaxies becomes incomplete in the range $9.5 \le \log(M/M_{\sun}) \le10$. This figure clearly shows how a difference between high- and low-density environments is present mainly for star-forming galaxies (except in the first redshift bin) up to $z \sim 1 - 1.5$ and generally not for quiescent galaxies. The ratio is $< 1$ for star-forming galaxies, and it is smaller in low-density environments by $\sim 0.2$ dex for environments measured with a fixed aperture radius of $R = 0.3$ Mpc. The difference between high- and low-density environments seems to get smaller increasing the fixed aperture radius. Nevertheless, an indication of low-mass star-forming galaxies being more present in low-density environments is visible in the data. Indeed, from the difference between the high- and low-density curves from the top panels, normalized to the sum in quadrature of the errors, it is possible to see how the difference between high- and low-density environments in the case of star-forming galaxies reaches values as high as $5\sigma$ in the $R = 0.3$ Mpc case, while being systematically larger than $3\sigma$. On the other hand, the difference between high- and low-density environments for quiescent galaxies is always in the $\pm 2\sigma$ range, being close to $3\sigma$ only in the $R = 2$ Mpc case. The monotonic trend with redshift of the intermediate-mass to low-mass ratio for the quiescent galaxies is an indication of a progressive steepening of the low-mass end of the quiescent GSMF with redshift (this seems to be at variance with what hinted by \cite{2016A&A...586A..23D}, although the different redshift range explored, mass completeness limit and environmental definition prevent us from drawing any firm conclusion from the comparison). Nevertheless, this monotonic trend with cosmic time is not observed for the star-forming galaxy population, and this can again be related to the gradual build-up of galaxy mass with cosmic time, in a complementary way than what found before with the high- to intermediate-mass end ratios. The low-mass end of the quiescent GSMF is gradually enhanced as more low mass galaxies are quenched with cosmic time, while the low-mass end of the star-forming GSMF is continuously replenished by galaxies that increase their stellar mass through ongoing star-formation activity. This result is in agreement also with what found by \citet[][see their Fig. 14]{2010A&A...523A..13P} and \citet[][see their Fig. 6]{2013A&A...556A..55I} as well as by many other works \citep[see \textit{e.g.}][and references therein]{2016A&A...590A.103M,2013ApJ...777...18M,2015MNRAS.447....2M,2014ApJ...783...85T}.

\subsection{The relative importance of quiescent and star-forming GSMF in different environments}
\begin{figure}
\centering
\resizebox{\hsize}{!}{\includegraphics{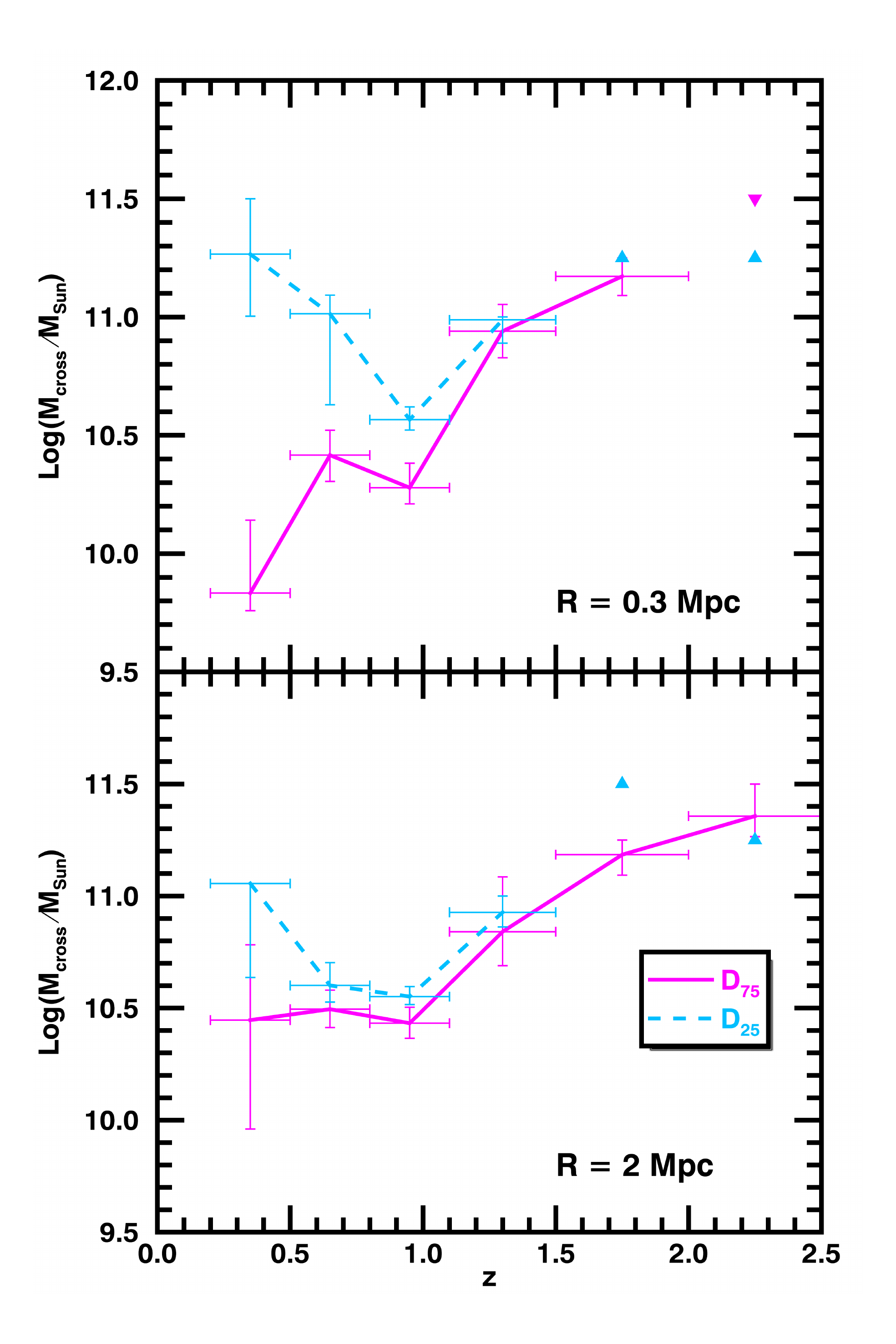}}
\caption{\textit{$M_{\rm cross}$}. Mass at which the star-forming and the quiescent GSMF intersect. The solid magenta line refers to high-density environments, while the dashed cyan line refers to low-density environments. Triangles (upward for low-density environments and downward for high-density environments) are upper limits to the value of $M_{\rm cross}$ for when the two GSMF do not intersect, defined as the mass at which the two GSMF are closer to each other. Galaxies in high-density environments undergo an accelerated evolutionary process with respect to galaxies in low-density environments.}
\label{mcross}
\end{figure}

Figure \ref{mcross} shows the mass at which the quiescent and the star-forming GSMF intersect ($M_{\rm cross}$) as a function of redshift and environment (\textit{i.e.} the mass above which the GSMF is dominated by the quiescent population). As for Fig. \ref{massratios}, the analysis has been limited at redshift $z \sim 2.5$ as at higher redshift the size of the quiescent galaxy sample becomes too limited. It can be seen how $M_{\rm cross}$ is higher in low-density environments compared to high-density ones up to redshift $z \sim 1.5$, where the two curves become indistinguishable. This is in agreement with the current paradigm of galaxy evolution, which predicts that massive galaxies became quiescent at earlier times compared to less massive galaxies. Therefore, as redshift increases, the mass at which the quiescent GSMF starts to dominate over the star-forming GSMF increases as well. The fact that $M_{\rm cross}$ is higher in low-density environments compared to high-density ones is an evidence of the fact that the processes that lead to the quenching of the star-formation and to the transformation of star-forming galaxies into quiescent galaxies are more efficient in high-density environments, leading to less massive galaxies being already quenched, while at the same redshift, in low-density environments, they will still be star-forming. In high-density environments $M_{\rm cross}$ is a monotonically increasing function of redshift, increasing from $\sim 10^{10} M_{\sun}$ at $z \sim 0.5$ to $10^{11.5}$ at $z \sim 2$. If we consider also upper limits to the value of $M_{\rm cross}$ derived when GSMF do not intersect, then an increase of $M_{\rm cross}$ as a function of redshift is roughly true also for low-density environments for redshifts $z \gtrsim 1$, while at lower redshifts Figure \ref{mcross} shows an upturn in the value of $M_{\rm cross}$. This upturn seems to become less evident going from $R = 0.3$ Mpc to $R = 2$ Mpc. The upturn at low redshifts of the $M_{\rm cross}$ in low-density environments is probably due to the lowest density environments probed by the fixed aperture method, especially on small scales (\textit{e.g.} $R = 0.3$ Mpc). In such underdense environments quiescent galaxies are rare, as the fraction of quiescent galaxies is higher in high-density environments (see \textit{e.g.} Fig. \ref{ETGFraction}). Therefore, the quiescent and star-forming GSMF will be more separated, especially at high masses, and the $M_{\rm cross}$ results higher.

\section{Comparison with previous studies}
\label{comparison}
Thanks to its excellent combination of multiwavelength coverage, fairly large area, and presence of datasets with a high statistical power, the UltraVISTA-COSMOS is a perfect field where to perform studies of galaxy evolution. For this reason it has been deeply exploited in several works. As the GSMF is a very powerful tool to study the galaxy formation from a statistical point of view, many studies have investigated its relation to galaxy environment. In this section we review some of the main works that studied the GSMF in different environments from low redshifts, using spectroscopic samples, up to high redshift, using photometric redshift surveys. 

The main spectroscopic survey that has been performed in the COSMOS field is the zCOSMOS Survey \citep[see][]{2007ApJS..172...70L}. Using the 10k spectroscopic sample of \citet{2009ApJS..184..218L} in synergy with the COSMOS photometric sample \citep{2007ApJS..172...99C,2010ApJ...708..202M} and the environmental estimate by \citet{2010ApJ...708..505K}, \citet{2010A&A...524A..76B} performed a thorough study of the GSMF in different environments up to $z = 1$. In their work, they found a difference between the GSMF of high- and low-density environments, with the massive end of the GSMF being more enhanced in high-density environments. This is qualitatively in agreement with our results in the common redshift range. In Figs. \ref{MFquiescent} and \ref{MFstarforming} we show a comparison between our GSMF and those of \citet[][see their figure 5]{2010A&A...524A..76B} for two common redshift bins. We compared quiescent and star-forming GSMF both in high- and low-density environments. GSMF have been normalized so to be equal at a given mass (that of the lowest mass bins considered for VIPERS GSMF, see below), which allows us to compare their shape. With the exception of quiescent galaxies in the range $0.8 \le z \le 1.1$ (where the GSMF of both high- and low- density environments are in good agreement), the GSMF of \citet{2010A&A...524A..76B} and those of our work show a slightly different shape, with the GSMF of \citet{2010A&A...524A..76B} displaying a steeper slope in the high-mass and/or low-mass end. 

The steeper slope of the zCOSMOS mass functions could be due to the different environment estimator used in our work (fixed aperture with $R = 0.3 - 2$ Mpc) and in \citet{2010A&A...524A..76B} (distance to the 5th nearest neighbour), to the different definition of quiescent and star-forming galaxies (photometric type coming from SED fitting estimate for \citet{2010A&A...524A..76B} \textit{vs} restframe color-color classification for our work) or to the fact that \citet{2010A&A...524A..76B} use spectroscopic redshift while we use photometric redshifts. Thanks to the higher statistics and lower mass limit compared to \citet{2010A&A...524A..76B}, we see a difference between high- and low-density environments for both passive and star-forming galaxies. 

A difference between our work and \citet{2010A&A...524A..76B} is confirmed also by the fraction of quiescent galaxies in different environments, shown in Fig. \ref{ETGFraction}. In particular, while the fraction of quiescent galaxies in both environments is comparable to the one derived in our work for the lowest mass bins considered by \citet{2010A&A...524A..76B}, their fractions become quickly higher than ours with increasing mass and especially in low-density environments. This is likely due to the different environmental estimator used in our work and in \citet{2010A&A...524A..76B} and it is likely the origin of the discrepancy observed in the values of $M_{\rm cross}$ between our work and \citet[][see their figure 7]{2010A&A...524A..76B}. In fact, while the value of $M_{\rm cross}$ in high-density environments is in agreement between our work and \citet{2010A&A...524A..76B}, we find an upturn in the values of $M_{\rm cross}$ in low-density environments which is totally absent in the work of \citet{2010A&A...524A..76B}. 

Although a quantitative comparison is difficult, the range of densities corresponding to low-density environments explored by our work is much lower than that in \citet{2010A&A...524A..76B}. Therefore, in such environments the quiescent galaxy population will be more under-represented and the corresponding quiescent GSMF will result depressed with respect to the star-forming GSMF, with the corresponding value of $M_{\rm cross}$ increased, as confirmed by the different fraction of quiescent galaxies.

In Figs. \ref{MFquiescent} and \ref{MFstarforming} we also report a comparison between our mass functions and those of \citet[][see their figure 4]{2016A&A...586A..23D}. Although with a different spectroscopic dataset \citep[VIPERS Survey,][]{2014A&A...566A.108G,2014A&A...562A..23G}, \citet{2016A&A...586A..23D} performed a thorough study of the GSMF in different environments at $z \lesssim 1$, finding consistent results with \citet{2010A&A...524A..76B}. Although their mass completeness limit allowed only a characterization of the high-mass end of the GSMF, still their GSMF for quiescent and star-forming galaxies in different environments are consistent within errorbars with those of our work in the overlapping redshift bins.

As the UltraVISTA sample offers high-quality photometric redshifts for a large statistical sample, some works have been performed at redshift $z > 1$ in the COSMOS field. A couple of recent works have explored the dependence of the GSMF on the environment using the same sample as we did in this work. Both \citet{2013ApJS..206....3S} and \citet{2015ApJ...805..121D} used a 2D Voronoi tessellation performed in subsequent redshift slices to study the environmental effects on the galaxy population and the GSMF. In particular, \citet{2015ApJ...805..121D} found a strong evidence for massive ($M > 10^{11} M_{\sun}$), quiescent galaxies showing an increasingly important difference between high- and low-density environments at $z \lesssim 1.5$. They found that the number density of massive quiescent galaxies in high-density environments is $\sim 10$ times higher than in low-density environments at redshift $z \lesssim 0.5$ (see their Figure 10). This is in agreement with the results of this work, which see environmental effects disappear for quiescent galaxies at $z \sim 2$, with the ratio of the high-mass end to the intermediate-mass end of the GSMF for quiescent galaxies being different by more than $3 \sigma$ in high-density environments with respect to low-density environments at $z \lesssim 0.5$ (see Figure \ref{massratios}). Both the work by \citet{2015ApJ...805..121D} and this work found no environmental effect for massive star-forming galaxies at any redshift. However, with our work we have been able also to extend the analysis to low-mass star-forming galaxies, finding an environmental effect up to $z \lesssim 1.5$.

It is also important to mention that our work is in agreement with what found in the UKIDSS UDS field \citep[see][]{2015MNRAS.447....2M}. By using the UKIDSS UDS dataset and the CANDELS Survey \citep[photometric redshifts,][]{2013ApJS..206...10G,2013ApJS..207...24G}, \citet{2015MNRAS.447....2M} found that the GSMF is different in high- and low-density environments up to $z \sim 1.5$. We show a comparison between our work and the work by \citet{2015MNRAS.447....2M} in three high-redshift bins in Fig. \ref{MFall}. GSMF are again normalized to be equal at $M = 10^{11} M_{\sun}$, so to be able to compare their shape in a consistent way. It can be seen how our GSMF and those of \citet{2015MNRAS.447....2M} are in good agreement except for the last redshift bin ($2.0 < z < 2.5$), where they show a different shape, with the GSMF of \citet{2015MNRAS.447....2M} characterized by a steeper slope. This difference in shape at high redshift could be due to the different high- and low-density environment definition (25th and 75th percentile of the volume density distribution in our work, $1\sigma$ deviation from the mean of the density distribution in \citealt{2015MNRAS.447....2M}). 


These comparisons show how we have been able to exploit the excellent UltraVISTA data set to extend previous works done at low redshift with spectroscopic surveys \citep[\textit{e.g.}][]{2010A&A...524A..76B,2016A&A...586A..23D} and to complement other works performed at high redshift with photometric redshift surveys \citep[\textit{e.g.}][]{2015ApJ...805..121D,2015MNRAS.447....2M}.

\section{Discussion}
\label{discussion}
Galaxies evolve in parallel with cosmic structures. As galaxies form, so do galaxy clusters, groups and the Large-Scale Structure (LSS) and transformations in galaxy properties happen at the same time as changes in their local and global environment. It is therefore expected some correlation between galaxy environment and galaxy properties, as a function of redshift. The current understanding of the effect of the environment on galaxy evolution is that environment plays a role in determining the cease of star formation in galaxies and in causing their transformation from blue, actively star-forming, disc-like objects to red, quiescent spheroidal systems.

This picture of galaxy formation in relation to environmental effects is supported by evidence both on the theoretical and the observational sides. For example, many mechanisms correlated to galaxy environment have been proposed to end the star formation in a galaxy \citep[see for example Figure 10 of][]{2003ApJ...591...53T} and many correlations have been found between the main observables and the density field (\textit{e.g.} galaxy color, mass, AGN activity, star-formation, morphology). Moreover, some works \cite[\textit{e.g.}][]{2010ApJ...721..193P} have proposed a separability of the processes that lead a galaxy to quiescence on the basis of mass and environment. In this work we present observational evidence of the presence of a complex interplay among galaxy mass, star-formation activity (or lack thereof) and local environment.

Although relying only on photometric redshifts, this work is able to recover with good accuracy the environmental trends of the GSMF, by making use of a method that has been fully tested on mock galaxy catalogues. In fact, in \citet{2016A&A...585A.116M} we carefully tested the effect that a density field measured using photometric redshifts has on the analysis of the GSMF in different environments, finding that no spurious effects are introduced and that the differences that are found would be greater if more precise redshifts were used. This is indeed an important argument in the present analysis: we can expect all the results that are described in this work to be more evident if spectroscopic redshifts were used. Of course, larger statistical samples at higher redshifts with precise redshift measurement are vital to perform this kind of studies. Here we discuss the results of Sect. \ref{results} in a theoretical framework of galaxy evolution, after having decisively ruled out possible effects due to photometric redshift uncertainties (see \citealt{2016A&A...585A.116M} and Sect. \ref{errorigrandiexplained} of this work).

We find that the galaxy population is different in different environments. High-density environments are populated by a higher fraction of quiescent galaxies and this distinction is particularly visible at high masses ($M \gtrsim 10^{11} M_{\sun}$) up to redshift $z \sim 2$. The lack of differences between the fraction of quiescent galaxies in different environments at higher redshifts is probably due to the fact that at higher redshifts structures are at an earlier stage of formation \citep[see \textit{e.g.}][]{2013ApJ...779..127C} and quiescent galaxies, even the massive ones, are rarer (while star-forming galaxies still dominate). This evidence is complemented by the total GSMF divided according to local environment, which shows how in high-density environments massive galaxies are more represented up to $z \sim 2$. Therefore, peaks in the density field seem to constitute a particular kind of environment where galaxies are more massive and more quiescent. Environment, therefore, plays a role in shaping the galaxy population and is connected to the build up of galaxy mass and to the end of the star formation. Environmental effects are visible since $z \sim 2$ and on scales of $R = 2$ Mpc, being effective for a long period of galaxy formation in a strong way.

A particularly interesting scenario for galaxy evolution is the one proposed by \citet{2015MNRAS.447..374G}. In their work, the authors used numerical simulations to investigate the new unified quenching model that they propose. In this model, a galaxy is quenched once the gas in its host halo becomes hot ($T \ge 10^{5.4}$ K) and this happens when the host halo reaches a mass of $10^{12} M_{\sun}$ (roughly corresponding to a stellar mass of $10^{10.5} M_{\sun}$). In this scenario, both \textquotedblleft mass quenching\textquotedblright$\:$ and \textquotedblleft environmental quenching\textquotedblright$\:$ \citep{2010ApJ...721..193P} are seen as separate evidences of the same underlying quenching mechanism due to the presence of hot gas. This theoretical model can be used to give an interpretation of our results. In particular, according to this model, the galaxies populating the high-mass end of the GSMF ($M \ge 10^{11} M_{\sun}$) are being quenched because they live in hot gas dominated haloes. Although they are found also in low-density environments \citep[see Figure 6 of][]{2015MNRAS.447..374G}, massive haloes ($M \ge 10^{12} M_{\sun}$) are found preferentially in high-density environments. This is the cause of the difference between high and low-density environments seen in the GSMF of quiescent galaxies at masses $M \ge 10^{11} M_{\sun}$. 

This difference is not seen in the GSMF of star-forming galaxies because massive galaxies in high-density environments are quenched, therefore they are not included in the star-forming GSMF. This goes in the direction of diluting the signal of potential differences in the high-mass end of the GSMF of star-forming galaxies as a function of environment. Instead, a difference is visible at low masses, with low-mass star-forming galaxies being more present in low-density environments. This is due to the fact that these galaxies live in too low-mass haloes to develop a hot gas environment and be quenched. Nevertheless, those living in high-density environments can still be quenched as satellites of more massive galaxies that live in hot gas dominated haloes. Therefore, the low-mass end of the star-forming GSMF is depleted in high-density environments compared to low-density ones. Interestingly, this trend should reflect in a difference in the low-mass end of the GSMF of quiescent galaxies in high-density environments, which seem to be absent in our data. This lack of a difference between the high- and low-density, low-mass end of the quiescent GSMF could be due to uncertainties in the photometric redshift calculation or in the distinction between quiescent and star-forming galaxies using the color-color diagram. Star-forming galaxies being the majority of the sample, a difference in the low-mass end of the GSMF can be recovered for them, but not for quiescent galaxies, which may suffer from residual contamination from star-forming galaxies at low masses. Nevertheless, a more accurate analysis, with more precise redshifts and a larger dataset has to be performed to solve the problem.

\section{Conclusions and summary}
\label{conclusions}
In this work we have used the GSMF and the high-precision photometric redshifts of the UltraVISTA Survey \citep{2012A&A...544A.156M,2013A&A...556A..55I} to outline a history of the role of the environment in galaxy evolution from $z = 3$ to $z = 0$. Thanks to the high-precision of the photometric redshift data-set used for this work, together with the accurate preparatory analysis performed in \citet{2016A&A...585A.116M}, the study described in this paper presents an additional perspective, both extending and complementing known results in the literature in a consistent way over a broader redshift range. Although derived with photometric redshifts the results presented in this work are robust and provide a reliable observational evidence to support theoretical scenarios of galaxy formation. Our main findings can be summarized as follows:

\begin{enumerate}
\item
The fraction of massive quiescent galaxies is higher in high-density environments compared to low density ones. The difference is visible up to a scale of $R = 2$ Mpc, and it is present up to redshift $z \sim 2$. The fraction of quiescent galaxies increases with mass and decreases with redshift.

\item
The shape of the galaxy stellar mass function is different in high-density and low-density environments for the total galaxy population. The high-mass end of the mass function ($\log(M^{\ast}/M_{\sun}) \in [11,11.5]$) is enhanced with respect to the intermediate-mass end ($\log(M^{\ast}/M_{\sun}) \in [10,10.5]$ in high-density environments) up to $z \sim 2$.

\item
The difference in the shape of the GSMF between high-density and low-density environments is visible for quiescent galaxies up to $z \sim 2$ and at masses $M > 10^{11} M_{\sun}$.

\item
The difference in the shape of the GSMF between high-density and low-density environments is visible for star-forming galaxies up to $z \sim 1.5$ and at masses $M < 10^{11} M_{\sun}$.

\item
No environmental effects seem to be visible in our data at $z \gtrsim 2$.

\item
The mass at which galaxies become quiescent at a given redshift is lower in high-density environments compared to low-density ones. This effect is visible up to redshift $z \sim 1.5$. In high-density environments, the mass at which the quiescent GSMF starts to dominate over the star-forming GSMF is a monotonically increasing function of redshift.
\end{enumerate}

We have shown that local environment plays indeed a role in shaping galaxy evolution, in the redshift range $0 \le z \le 2$. High-density environments show an enhanced fraction of massive ($\sim 10^{11} M_{\sun}$) quiescent galaxies, compared to low-density ones. At high redshift many structures may be at an earlier stage of formation, with clear environmental dependencies not yet in place. Although present at $z > 2$, structures with a clear segregation of quiescent galaxies may be rare, requiring a larger area than the COSMOS field to be detected in a sufficient number. As at these redshifts also lower number statistics and larger uncertainties in the photometric redshift determination may start to affect the sample, these results would benefit from a confirmation by the means of wide-area spectroscopic redshift surveys such as Euclid and WFIRST.

\section*{Aknowledgements}
NM wishes to thank Micol Bolzonella, Iary Davidzon, and Alice Mortlock for providing data points and suggestions that helped to significantly improve the paper. NM also wishes to thank St\'{e}phane Arnouts for the many useful discussions regarding this work. This work is based on data products from observations made with ESO Telescopes at the La Silla Paranal Observatory under ESO programme ID 179.A-2005 and on data products produced by TERAPIX and the Cambridge Astronomy Survey Unit on behalf of the UltraVISTA consortium. We acknowledge the financial contributions by grants ASI/INAF I/023/12/0, PRIN MIUR 2010-2011 \textquotedblleft The dark Universe and the cosmic evolution of baryons: from current surveys to Euclid\textquotedblright, and PRIN MIUR 2015 \textquotedblleft Cosmology and Fundamental Physics: illuminating the Dark Universe with Euclid\textquotedblright. 

\bibliographystyle{mnras}
\bibliography{Malavasi3mnras}

\begin{appendix}
\section{A test on the effect of photometric redshift uncertainties}
\label{errorigrandiexplained}

\begin{figure*}
\centering
\resizebox{\hsize}{!}{\includegraphics{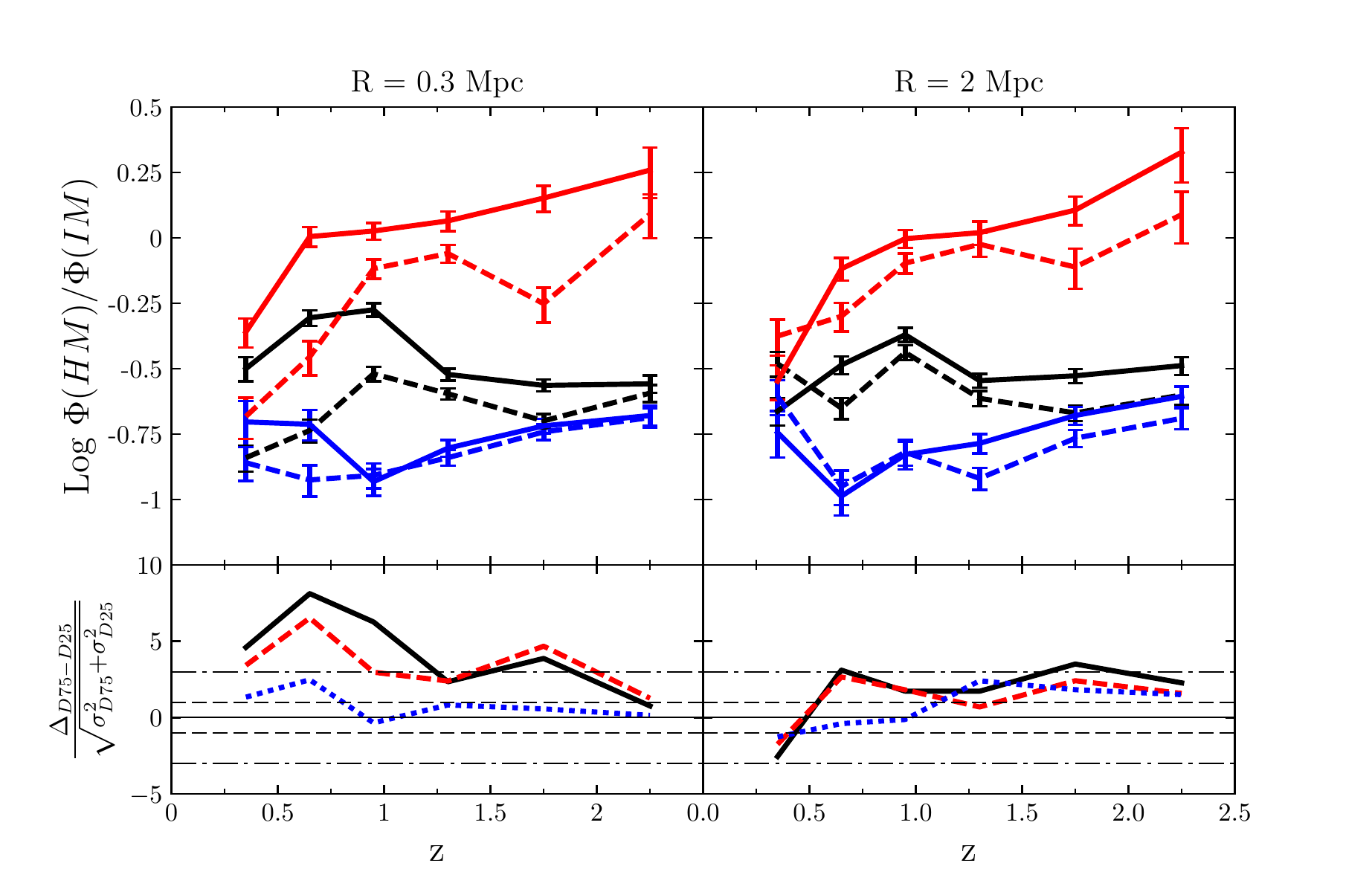}}
\caption{\textit{Shape of the GSMF - increased errors}. As Figure \ref{massratios}, but with larger photometric redshift errors at high redshift (see text). Increased errors for photometric redshifts at high-redshift do not affect the results of this work.}
\label{massratioslowzhighz}
\end{figure*}

As discussed in \citet{2016A&A...585A.116M}, photometric redshift uncertainties are a major limitation in the reconstruction of the density field. Nevertheless, in \citet{2016A&A...585A.116M} we demonstrated that it is still possible to perform a study of the GSMF in different environments using photometric redshifts provided that their uncertainty is small ($\sigma_{\varDelta z/(1+z)} \lesssim 0.01$). In this case, differences between high- and low-density environments that are present in the GSMF calculated using each galaxy's true redshift up to $z \sim 2.5$ result damped when using photometric redshifts, but they will still be recovered. Following what shown in \citet{2016A&A...585A.116M} and as explained in Sect. \ref{data}, we have chosen for this work an uncertainty value for the photometric redshifts of $\sigma_{\varDelta z/(1+z)} = 0.01$ which may be, nevertheless, underestimated at high redshift. Moreover, the photometric redshift uncertainty depends on $K_S$-band magnitude, as shown, for example, in Fig. 2 \citet{2013ApJS..206....3S}. This figure shows the photometric redshift uncertainty as a function of $K_S$-band magnitude and redshift, together with the median $K_S$-band magnitude of a sample of galaxies extracted from the UltraVISTA Survey and similar to the one used in this work. For this reason, we have tested also the effect of a larger uncertainty. Following Fig. 2 of \citet{2013ApJS..206....3S} we have redone our work assuming a photometric redshift uncertainty of
\begin{equation}\label{errorigrandi}
\sigma_{\varDelta z/(1+z)} =
\begin{cases}
0.01 & \text{for} \: z \le 1.5 \\
0.03 & \text{for} \: z > 1.5 \\
\end{cases}
\end{equation}
As the main purpose of this paper is to study the shape of the GSMF in different environments, we tested whether the differences that we see between the GSMF in high- and low-density environments (Figs. \ref{massratios} and \ref{massratioslmlz}) are maintained when considering larger errors at higher redshift. Because in the case of the ratio between the intermediate-mass and low-mass end of the GSMF we limit the analysis at $z \le 1.5$, we only report for comparison in Fig. \ref{massratioslowzhighz} the ratio of the high-mass to the intermediate-mass end of the GSMF (see Fig. \ref{massratios}), performed with the higher photometric redshift uncertainty value at high redshift. It can be seen how, even with larger photometric redshift errors, the trends are maintained, without significant differences. The increase in the redshift error affects only the analysis at $z \ge 1.5$ and in a negligible way.

\end{appendix}

\bsp	
\label{lastpage}
\end{document}